\documentclass[11pt]{article}
\pdfoutput=1
\usepackage{subfigure}
\usepackage{amsmath}
\usepackage{amsthm}
\usepackage{algorithmic}
\usepackage{algorithm}
\usepackage{fullpage}
\usepackage{xspace}
\usepackage{url}
\usepackage{graphicx} 

\theoremstyle{remark}

\newcommand{\ou}{Ornstein-Uhlenbeck\xspace}
\newcommand{\Var}{\mbox{Var}}

\begin{document}

\title{
Financial LPPL Bubbles with Mean-Reverting Noise in the Frequency Domain
}

\author{
Vincenzo Liberatore 
\thanks{Division of Computer Science,
Case Western Reserve University,
10900 Euclid Avenue,
Cleveland, Ohio 44106-7071, USA.
E-mail: {\tt vl@case.edu\/}.
URL: {\tt http://vincenzo.liberatore.org/\/}. }
}

\maketitle

\begin{abstract}
The log-periodic power law (LPPL) is a model of asset prices during 
endogenous bubbles.
A major open issue is to verify the presence of LPPL in price sequences
and to estimate the LPPL parameters.
Estimation is complicated by the fact that daily LPPL returns are typically 
orders of magnitude smaller than measured price returns, suggesting that 
noise obscures the underlying LPPL dynamics.
However, if noise is mean-reverting, it would quickly cancel out
over subsequent measurements. 
In this paper, we attempt to reject mean-reverting noise from price sequences 
by exploiting frequency-domain properties of LPPL and of mean reversion.
First, we calculate the spectrum of mean-reverting \ou noise and devise
estimators for the noise's parameters.
Then, we derive the LPPL spectrum by breaking it down into its 
two main characteristics of power law and of log-periodicity.
We compare price spectra with noise spectra during historical 
bubbles.
In general, noise was strong also at low frequencies and,
even if LPPL underlied price dynamics, LPPL would be obscured by noise.
\end{abstract}

\section{Introduction}
Financial bubbles and busts have devastating effect on the economy
and on markets.
However, the existence and characteristics of bubbles are notoriously hard
to ascertain if not with hindsight. 
This paper contributes to the investigation of financial bubbles 
within the LPPL framework \cite{lppl-book}, and 
specifically it examines its frequency-domain properties
under a mean-reverting noise model.

Endogenous financial bubbles have been modeled as a 
{\em log-periodic power law (LPPL)\/} \cite{lppl-book}. 
The LPPL model has two main characteristics:
\begin{itemize}
\item Super-exponential growth, leading to a {\em critical time\/} at which
the asset price will burst ({\em power law\/}), and
\item Oscillations that become progressively faster as the critical
time approaches ({\em log-periodicity\/}).
\end{itemize} 
Super-exponential growth is a sign that price growth is unsustainable.
The oscillatory behavior indicates an incipient system failure,
and is often associated with increasingly more rapid and pronounced swings 
in investor sentiment \cite{biased-diffusion}.

Previous work has paid considerable attention to fitting the LPPL law
to historical time series of financial bubbles, and
a recent summary reviews the state of the art
\cite{lppl-fit}.
Current methods for LPPL bubble detection have been 
tested in an on-line experiment \cite{sornette-gamble}.
Previous work mostly focus on the two related issues of 
statistical significance and of noise in the data.
The LPPL model parameters can be fit to the price sequences 
via non-linear least-squares \cite{lppl-fit}, and
we have previously proposed efficient parallel algorithms for least-square
fitting \cite{vl-lppl}.
The least-square algorithm returns an estimate for the LPPL parameters
as well as a residual error. 
However, least squares always estimate parameters for the LPPL model, 
regardless of whether LPPL underlies actual price dynamics or not.
The least square residual error gives a sense of the presence of LPPL: 
if the final mean squared error is significantly smaller than the error at
the beginning of the numerical fit, 
then there is some evidence for an underlying LPPL process.
However, reliance on mean squared errors can produce spuriously 
high goodness of fit \cite{lsq-spurious74, lsq-spurious86}.
Although LPPL is often visible intuitively in price series and 
non-linear least squares substantially reduce the mean squared error over
an initial exponential fit, 
the log-periodic component of the S\&P 500 is not 
statistically significant prior to the 1987 crash if the last year of data is 
removed \cite{F01a}.
Furthermore, Bayesian methods fundamentally reject the hypothesis 
that LPPL underlies price dynamics during a bubble \cite{CF06a}.
An alternative approach is to verify the non-linear least squares
using unrelated tests, such as bounding the range of acceptable LPPL
parameters or using a Lomb transform \cite {lppl-fit}.
However, log-periodic variations prior to large drawdowns fail to satisfy the
parameter restrictions in the LPPL model \cite{regime-switch}.

In fitting LPPL to prices, 
the fundamental problem is that the daily LPPL returns are 
typically orders of magnitude smaller than the measured price returns.
In other words, a hypothetical LPPL signal would be swamped by overwhelming 
noise, and thus it is difficult to ascertain its presence and its parameter 
values.
Furthermore, it is also possible that the noise variance increases over time,
in which case the observed prices tend to progressively differ more 
significantly from an underlying deterministic LPPL.
When noise variance increases, the noise effectively becomes the signal.
As such, LPPL estimation is plagued by a low signal-to-noise
ratio, which additionally may be decreasing over time.
However, 
an LPPL model with mean-reverting noise (e.g., \ou)
has been recently proposed \cite{OU}.
If noise is mean-reverting, it has bounded variance, and thus the measured
prices fits more tightly around the underlying LPPL. 
Furthermore, although a mean-reverting process may be characterized by large 
daily return, the daily gyrations will rapidly cancel each other out.
In other words, mean-reverting noise is characterized by relatively 
low power at low frequencies. 
Thus, mean-reversion would make it possible to reconstruct the underlying
LPPL by focusing on the lower frequency components of measured prices.

In this paper, we examine LPPL with mean-reverting \ou noise in the 
frequency domain.
The paper investigates the mean-reverting properties of the noise, and
whether the de-noised power spectrum shows a deterministic LPPL signature.
Section \ref{sec:lppl} describes the LPPL model.
Section \ref{sec:ou} gives the necessary background on \ou processes, 
analyzes its spectrum, and discusses two methods for estimating its 
parameters.
Section \ref{sec:reflect} gives our methodology for calculating spectra
of finite price sequences.
Section \ref{sec:lpplspectrum} discusses the LPPL spectrum by breaking it
down into its two main features of power law and log-periodicity.
Section \ref{sec:de-noise} discusses noise rejection in noisy LPPL price
sequences.
Section \ref{sec:evaluation} evaluates the approach on major historical bubbles.
Section \ref{sec:conclusion} concludes the paper and outlines possible 
future work.

\section{LPPL}
\label{sec:lppl}
\begin{figure}
\begin{center}
\includegraphics[width=8cm]{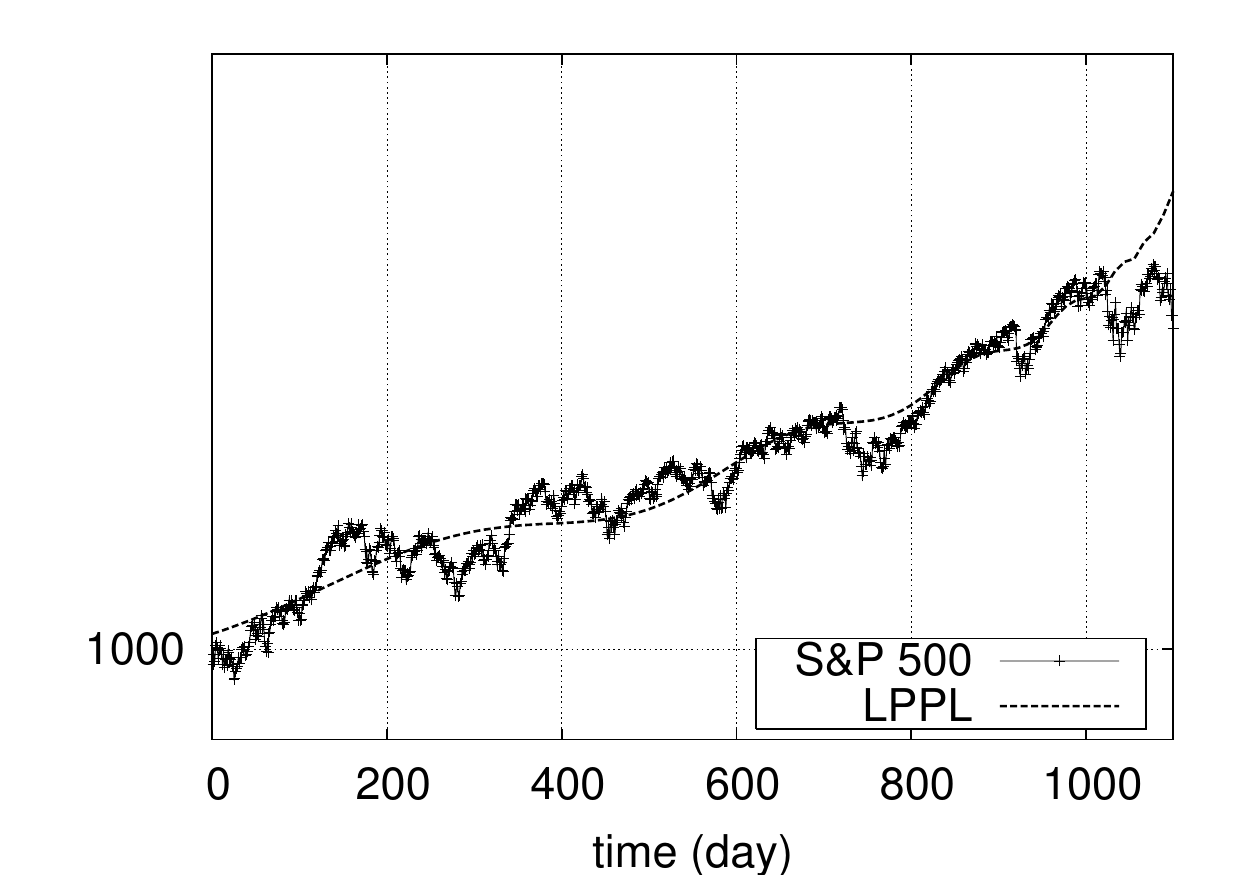}
\end{center}
\caption{Log-scale S\&P 500 and an LPPL fit from July 2003 to June 2007.}
\label{fig:lppl}
\end{figure}

The {\em log-periodic power law\/} is a function:
\begin{equation}
\label{eq:lppl}
\ell(t) = A - B (T - t)^m (1 + C \cos (\omega \ln (T - t) + \phi))\;,
\end{equation}
where $B > 0$ and $0 < m \leq 1$. 
The LPPL function is a model for a sequence of price logarithms 
$p(0), p(1), \dots , p(n-1)$ in the sense that $p(i) = \ell(i) + \nu(i)$,
where $\nu(i)$ is random noise independent of $\ell$.
If we assume that $\ell(T) = A$, 
the LPPL function $\ell$ is defined for $t \leq T$.
In particular, since $\ell$ is a model for a finite sequence of prices,
we will also restrict $\ell$ over $t \geq 0$.
As matter of notational convention, we remark that we call price 
sequence and denote as $p(0), p(1), \dots , p(n-1)$ what in reality
is the natural logarithm of prices. 
The convention makes our notation more compact, but it should be kept in
mind that, for example, our daily price returns are basically price ratios.
Figure \ref{fig:lppl} shows the S\&P 500 daily closing prices and a
least squares LPPL fit for the four years between July 2003 and June 2007.
If $m = 1, C = 0$, LPPL reduces to an exponential model
of the price sequence.
If $m < 1$, then LPPL grows super-exponentially until the
{\em critical time\/} $T$.
If $C, \omega > 0$, then LPPL exhibits oscillations that 
become progressively more frequent as $t$ approaches the
critical time. 
Since $\ell$ is a deterministic signal and $\nu$ is a wide-sense
stationary process independent of $\ell$,
most properties of stationary signals can be applied \cite{P01a}.
Since $\nu$ is independent of $\ell$,
$|P(f)|^2 = |L(f)|^2 + S(f)$,
where $L(f)$ is the transform of $\ell$,
$P(f)$ is the transform of $p$, and 
$S(f)$ is the noise power.

\section{Ornstein-Uhlenbeck Noise}
\label{sec:ou}

\subsection{Background}
The {\em \ou process\/} $\nu(t)$ is the solution of the stochastic differential
equation:
$$\nu'(t) = - \frac{\nu(t)}{\tau} + \sigma \Gamma(t)\;,$$
where $\Gamma(t)$ is Gaussian white noise,
$\tau \geq 0$ is called the {\em relaxation  time\/}, and
$\sigma \geq 0$ is called the {\em diffusion constant\/} 
\cite{ou-survey}.
The \ou process can be interpreted as white noise that has been 
filtered by a first-order system with cut-off frequency 
$f_c = 1 / (2 \pi \tau)$.
When $\tau \rightarrow \infty$, the \ou process reduces to a Weiner process.
As $t \rightarrow \infty$, $\Var [ \nu(t) ] = \sigma^2 \tau / 2$
and $\mbox{Cov} [ \nu(t+h), \nu(t) ] = \sigma^2 \tau e^{-|h|/\tau} / 2$
\cite{ou-survey}.
A \ou process can be simulated by means of:
\begin{equation}
\label{eq:ousim}
\nu(t + 1) = \nu(t) e^{-1 / \tau} + 
             \sqrt{\frac{\sigma^2 \tau}{2} (1 - e^{-2/\tau})} u_t\;,
\end{equation}
where $u_t$ is a sample value of a normal random variable 
\cite{gillespie1991markov}.
The stationary \ou process can be simulated by using recurrence
\eqref{eq:ousim} starting from $\nu(0)$ being the sample of a Gaussian
random variable with variance equal to 
$\lim_{t \rightarrow \infty} \Var [ \nu(t) ] = \sigma^2 \tau / 2$.

\subsection{Spectrum}
\label{sec:ouspectrum}
\begin{figure}
\begin{center}
\includegraphics[width=8cm]{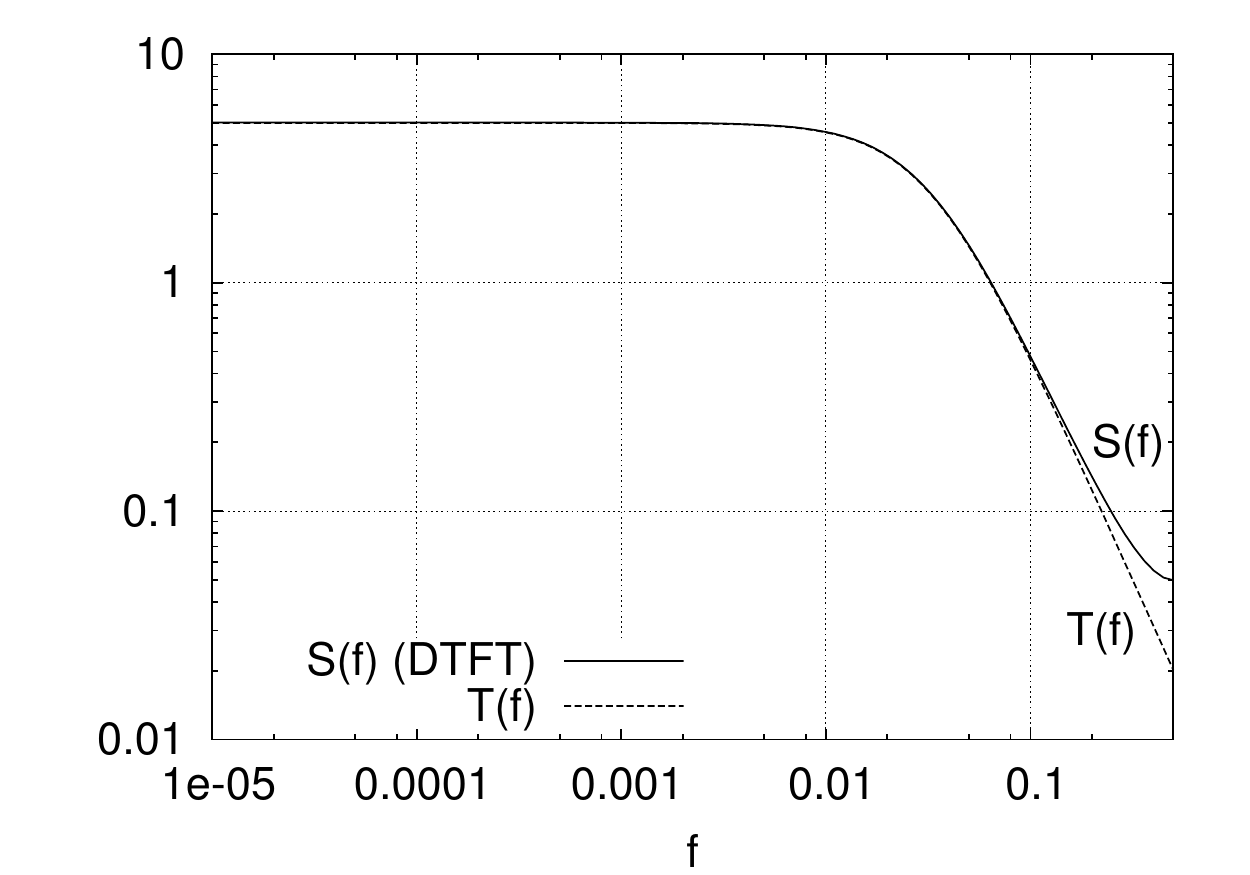}
\end{center}
\caption{Power spectrum of the Fourier and discrete-time Fourier transform
for $\tau = 5$, $\sigma = 1 / 5$.}
\label{fig:SvsT}
\end{figure}

The two-sided power spectrum is defined for $t \rightarrow \infty$, is equal
to \cite{ou-original}:
\begin{align*}
T(f) & = \frac{\sigma^2 \tau^2}{1 + 4 \pi^2 \tau^2 f^2} & &
\left( 0 \leq f \leq \frac{1}{2} \right)
\end{align*}
and it is the same as that of a first-order low-pass filter with cut-off 
frequency $f_c$.
Since we will be dealing with sampled \ou processes, it is also useful
to derive the spectrum of the discrete-time Fourier transform.
The transform of the auto-correlation function is
\begin{eqnarray*}
S(\omega) & = &
\sum_{h = - \infty}^{\infty} 
\frac{\sigma^2 \tau}{2} e^{-|h|/\tau} e^{-j \omega h} \\
& = &
\frac{\sigma^2 \tau}{2} 
\left(-1 + 
\sum_{h = 0}^{\infty} e^{-h/\tau} (e^{-j \omega h} + e^{j \omega h}) \right) \\
& = &
- \frac{\sigma^2 \tau}{2} + 
\sigma^2 \tau \sum_{h = 0}^{\infty} a^{h} \cos(\omega h) \;,
\end{eqnarray*}
where $a = e^{-1 / \tau}$.
Since, it is known that for $|r| < 1$ \cite[eq. 13.2(29)]{handbook}
$$\sum_{h = 0}^{\infty} r^h \cos (h k) = 
\frac{1 - r \cos k}{1 - 2 r \cos k + r^2}\;,$$
we have that
$$S(\omega) = 
\frac{\sigma^2 \tau}{2} 
\left( -1 + 2 \frac{1 - a \cos \omega}{1 - 2 a \cos \omega + a^2}\right) =
\frac{\sigma^2 \tau}{2} \frac{1 - a^2}{1 - 2 a \cos \omega + a^2}\;.$$
In terms of frequency $f = \omega / (2 \pi)$,
$$S(f) = 
\frac{\sigma^2 \tau}{2} \frac{1 - a^2}{1 - 2 a \cos (2 \pi f) + a^2}\;.$$
Figure \ref{fig:SvsT} shows the power of the
discrete-time Fourier transform $S(f)$ and of 
the continuous Fourier transform $T(f)$
for $\tau = 5$ and $\sigma = 1 / 5$.
The power spectra are similar except at the highest frequencies.
At low frequencies, since $a \simeq 1 - 1 / \tau$ and $a^2 \simeq 1 - 2 / \tau$,
$T(f) \simeq S(f) \simeq \sigma^2 / (1-a)^2 \simeq \sigma^2 \tau^2$.

\subsection{\ou Estimation}
We now turn to the estimation of the parameters $\sigma$ and $\tau$ in the
\ou noise. 
We will follow two approaches for parameter estimation. 
The first approach is based on the intuition that on short time frames
the noise dominates the LPPL signal (e.g., \cite{CF06a}).
Therefore, the noise parameters could be determined with high accuracy
by looking at diffusion and mean reversal over short intervals,
possibly corrected by a first-order trend estimate.
The second approach is pessimistic: it assumes $\tau \rightarrow \infty$
and finds the largest value of $\sigma$ that is consistent with the observed
signal. In other words, the pessimistic approach attempts to explain 
the observed signal as much as possible as Weiner noise.

\subsubsection{Maximum Likelihood}
High-frequency estimation is based on the assumption that over short time
scales the signal is completely dominated by noise.
In this case, $\ell(t+1) \simeq \ell(t)$, and since 
$p(t) = \ell(t) + \nu(t)$, we have that
$\nu(t+1) - \nu(t) \simeq p(t+1) - p(t)$.

Given a sequence of observations $\nu(0), \nu(1), \dots, \nu(n-1)$ 
of an \ou process, the maximum likelihood estimates of $\tau$ and $\sigma^2$
are 
$$\hat{\tau} = \ln \frac{S_{xx}}{S_{xy}}\;,$$
$$\hat{\sigma}^2 = \frac{2 \hat{\tau}}{1 - \hat{a}^2} \frac{1}{n - 1} 
(S_{yy} -2 \hat{a} S_{xy} + \hat{a}^2 S_{xx})\;,$$
where $\hat{a} = e^{-1/\hat{\tau}}$, 
$S_{xx} = \sum_{i = 1}^{n-1} \nu^2(i-1)$,
$S_{xy} = \sum_{i = 1}^{n-1} \nu(i-1) \nu(i)$, and
$S_{yy} = \sum_{i = 1}^{n-1} \nu^2(i)$ \cite{oucalibrate}.
Maximum likelihood estimation is an estimation technique that operates 
exclusively in the time-domain.

However, the $\nu$ values are only known indirectly through
the differences $\nu(t+1) - \nu(t)$.
Define $\alpha_t = \nu(t) - \nu(0)$ and observe that $\alpha_t$ is known
via the telescoping summation
$\nu(t) = \nu(0) + \sum_{i = 0}^{t-1} (\nu(i+1) - \nu(i))$.
Since the \ou process has no drift, $S_x S_{xy} = S_y S_{xx}$, where
$S_x = \sum_{i = 1}^{n-1} \nu(i-1)$ and 
$S_y = \sum_{i = 1}^{n-1} \nu(i)$ \cite{oucalibrate}.
Therefore,
\begin{eqnarray*} 
0 & = &
\sum_{i = 1}^{n-1} (\nu(0) + \alpha_{i-1}) 
\sum_{j = 1}^{n-1} (\nu(0) + \alpha_{j-1}) (\nu(0) + \alpha_{j}) -
\sum_{i = 1}^{n-1} (\nu(0) + \alpha_{i})
\sum_{j = 1}^{n-1} (\nu(0) + \alpha_{j-1})^2 \\
& = &
\sum_{i = 1}^{n-1} \sum_{j = 1}^{n-1} 
(\nu(0)^3 + \nu(0)^2 (\alpha_j + \alpha_{j-1} + \alpha_{i-1}) +
\nu(0) (\alpha_j \alpha_{j-1} + \alpha_{i-1} \alpha_j + 
\alpha_{i-1} \alpha_{j-1}) \\
& & 
+ \alpha_{i-1} \alpha_j \alpha_{j-1}) \\
& & - \sum_{i = 1}^{n-1} \sum_{j = 1}^{n-1} 
\nu(0)^3 + \nu(0)^2 (2 \alpha_{j-1} + \alpha_i) +
\nu(0) (\alpha_{j-1}^2 + 2 \alpha_i \alpha_{j-1}) + \alpha_i \alpha_{j-1}^2 \\
& = &
(n - 1) \nu(0) A_{xy} + \nu(0) A_x^2 + A_x A_{xy} - (n - 1) \nu(0) A_{xx} -
\nu(0) A_x A_y - A_y A_{xx}\;,
\end{eqnarray*}
where
$A_x = \sum_{i = 1}^{n-1} \alpha_{i-1}$,
$A_y = \sum_{i = 1}^{n-1} \alpha_i$,
$A_{xy} = \sum_{i = 1}^{n-1} \alpha_i \alpha_{i-1}$, and
$A_{yy} = \sum_{i = 1}^{n-1} \alpha_i \alpha_i$,
Solving for $\nu(0)$ now gives
$$\nu(0) = \frac{A_y A_{xx} - A_x A_{xy}}
{(n - 1) A_{xy} - (n - 1) A_{xx} + A_x^2 - A_y A_x}\;.$$
Given the estimate for $\nu(0)$ and the $\alpha_i$'s, the $\nu(i)$'s 
can be estimated, and thus $\tau$ and $\sigma^2$.

To increase the estimate accuracy in the presence of
an underlying deterministic trend $\ell$, 
we estimate a linear fit to the log-prices $p$ and subtract it from
$p(t)$.

\begin{figure}
\begin{center}
\includegraphics[width=8cm]{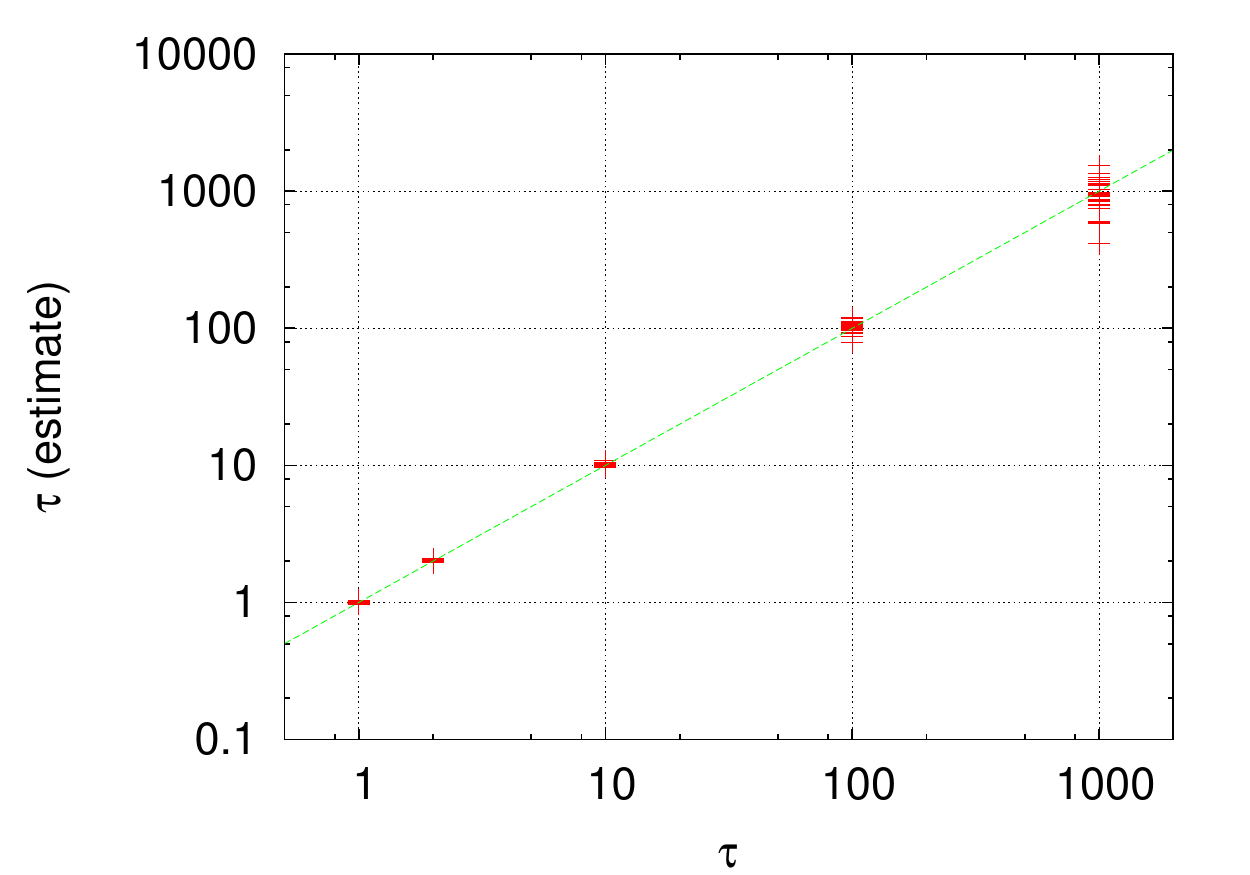}
\end{center}
\caption{Estimate $\hat{\tau}$ versus its underlying value $\tau$.
For each value of $\tau$, 30 Monte Carlo simulations were executed
over intervals of length $n = 25000$.}
\label{fig:tau-est}
\end{figure}

Monte Carlo simulations were conducted to ascertain the accuracy of the 
maximum likelihood estimation.
The estimation was quite accurate when the underlying LPPL trend 
gave negligible contributions to the $\alpha_i$'s, but less accurate
otherwise.
Additionally, even in the complete absence of an underlying trend,
the estimate $\hat{\tau}$ of $\tau$ became progressively less 
accurate as $\tau$ gets larger (Figure \ref{fig:tau-est}). 
An intuitive explanation is that when $\tau$ is large, then 
mean reversion affects only slightly the $\alpha_i$'s, making it hard 
to obtain an accurate fix on $\tau$.
In principle, the problem could be obviated by formulating an estimate
with returns over longer periods of time 
(i.e., $\tilde{\nu}(i) = \nu(k i), k > 1$).
However, the problem is compounded if an underlying trend $\ell$ 
has relatively large power at frequencies smaller than
the critical frequency $f_c = 1/ (2 \pi \tau)$.
In this case, the underlying trend $\ell$ affects the signal at the same
frequencies at which the relaxation term does, thus obscuring the exact 
contribution of the relaxation constant.
The case when $\tau$ is large will be addressed by pessimistic estimation.

\subsection{Pessimistic Estimation}
\label{sec:pessimistic}
In pessimistic estimation, we assume that the noise is generated by 
a Weiner process (i.e., $\tau \rightarrow \infty$) and 
estimate $\sigma$ using the constraint $|P|^2 = |L|^2 + S \geq S$
that the noise power should not exceed the observed signal power.
The estimate is pessimistic in that it assumes no mean reversion and thus
strong noise even at low frequencies.

Let $S_{\tau}(f)$ be the power spectrum of an \ou process when 
the relaxation constant is $\tau$ and define 
$S_{\infty}(f) = \lim_{\tau \rightarrow \infty} S_{\tau}(f)$.
Then, if $f > 0$, 
$$S_{\infty}(f) = \lim_{\tau \rightarrow \infty} S_{\tau}(f) =
\frac{\sigma^2}{4 (1 - \cos(2 \pi f))} 
\lim_{\tau \rightarrow \infty} \tau \left(1 - e^{-2 / \tau}\right) =
\frac{\sigma^2}{2 (1 - \cos(2 \pi f))}\;,$$
and
$$S_{\infty}(0) = 
\lim_{\tau \rightarrow \infty} \frac{\sigma^2 \tau}{2} \frac{1 - a^2}{(1 - a)^2}
= \sigma^2\;.$$
It should be the case that $P(f) \leq S_{\infty}(f)$ for all $f$'s.
In practice, however, the observed signal is the realization of a stochastic
signal, and so $P(f)$ may or may not exceed $S_{\infty}(f)$ \cite{loeve}.
Hence, we only impose integral constraints of the form 
$$\int_{f^-}^{f^+} S_{\infty}(f) df \leq \int_{f^-}^{f^+} P(f) df\;,$$
where $f^-$ and $f^+$ are parameters to be determined with
$0 < f^- < f^+ \leq 1/2$.
Furthermore, the spectrum $P$ is calculated only at the discrete points
$f_i = i / n$, so the constraint becomes
$$\sum_{i = h}^k S_{\infty}(f_i) \leq \sum_{i = h}^k P(f_i)\;,$$
which then leads to
\begin{equation}
\label{eq:pessimistic}
\sigma^2 \leq 
\frac{2 \sum_{i = h}^k P(f_i)}{\sum_{i = h}^k 1 / (1 - \cos (2 \pi f_i))}\;.
\end{equation}
A pessimistic estimate can be obtained by calculating the bounds
for $h_0 = 1$, $k_j = h_{j+1} = (1 + \alpha) h_j$ 
(for some constant $\alpha > 0$, e.g., $\alpha = 1$)
and then taking the most conservative bound.

The numerical calculation of the bounds is complicated by the fact that
the summations in \eqref{eq:pessimistic} involve terms whose value can differ
by orders of magnitude.
In this case, if the largest terms are added first, due to the finite
representation precision, the smallest terms are likely to be truncated.
In an extreme scenario, many small terms could collectively dominate the 
value of the summation but since they are added individually to a larger 
partial sum, their contribution would be lost.
In general, a summation is numerically more accurate when it always adds
terms of comparable magnitude.
Since $S_{\infty}$ is a decreasing function
(and the same holds in the expectation for $P$),
the numerical calculation should always start from the largest
frequency $f_k$ and proceed backward to $f_h$.
(As an aside, however, the numerical calculation is only needed for
the numerator, since the denominator could be replaced by an integral that is 
solvable analytically.)

\begin{figure}
\begin{center}
\includegraphics[width=8cm]{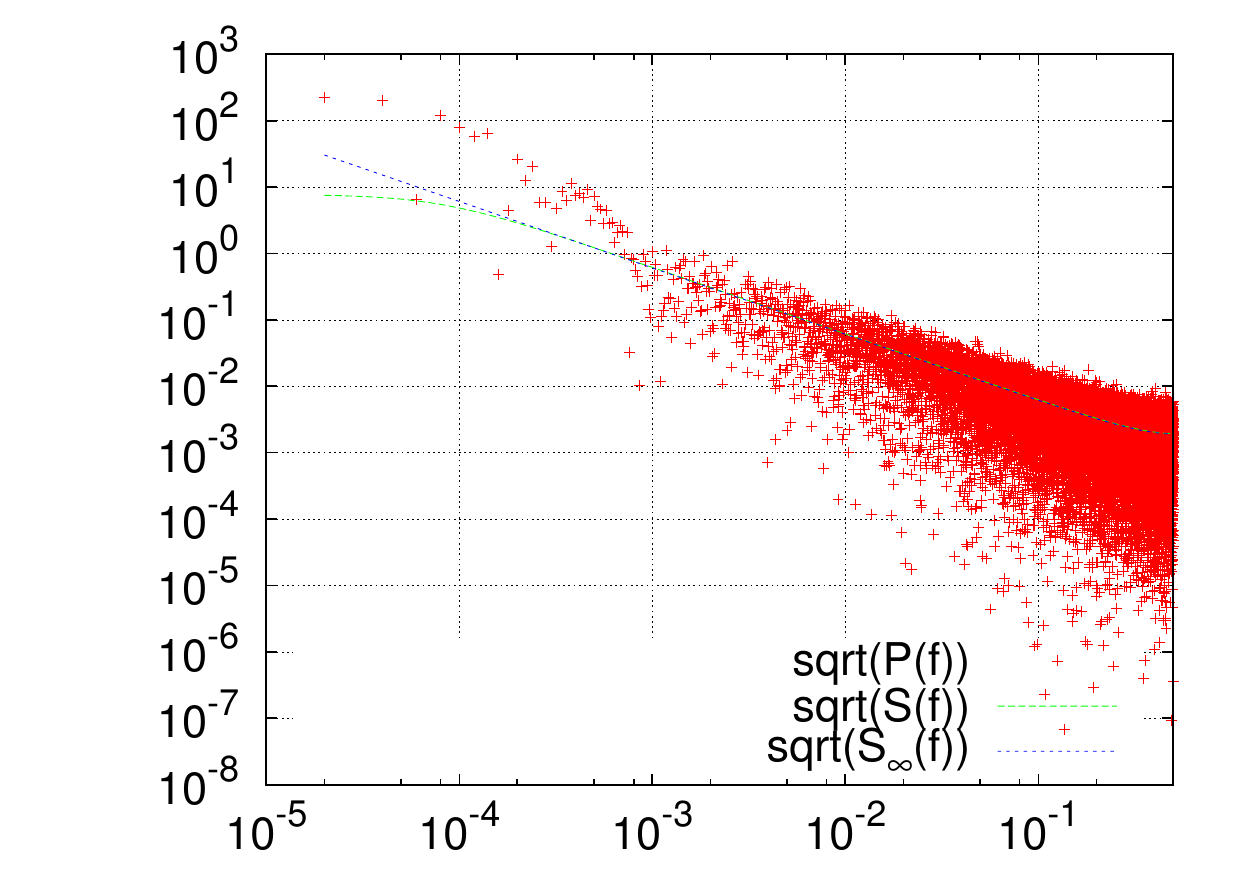}
\end{center}
\caption{DFT amplitude of a log-periodic function,
($A = 10, B = 0.008, T = 26000, m = 0.7, n = 25000, C = 0.4, \omega = 2 \pi, 
\phi = \pi$),
underlying \ou noise 
($\sigma^2 = .000015, \tau = 2000, f_c = 7.96 \cdot 10^{-5}$),
and pessimistic estimate ($\alpha = 1$, $\sigma^2 = .0000145$).}
\label{fig:pessimistic}
\end{figure}

Figure \ref{fig:pessimistic} shows an example of pessimistic estimate 
in a case when $\tau = 2000$ is relatively large and difficult 
to estimate because $P(f_c)$ is dominated by 
the power of the underlying signal $\ell$.
The pessimistic estimate was quite accurate, and $S_{\infty}$ differs from 
the underlying $S(f)$ only for the smallest frequency values.
Similar conclusions were supported by several additional Monte Carlo
simulations.

\section{Reflection}
\label{sec:reflect}
The \ou process describes an infinitely long signal.
On the contrary, $p(0), p(1), \dots, p(n - 1)$ is a finite sequence.
Similarly, $\ell$ is defined in the finite interval $[0, T]$.
In the calculation of the spectrum, an ambiguity arises 
regarding the behavior of $p$ and $\ell$ outside of their definition
interval.
In many applications, finite sequences are multiplied by a window function,
or they are implicitly assumed to be periodic.
Implicit periodization matches directly the discrete Fourier transform (DFT), 
in the sense that, given the sequence $p(0), p(1), \dots, p(n-1)$, 
the DFT assumes that the signal is infinite, periodic, with period $n$,
and that $p(i) = p(i - n)$ for $i = 0, \pm 1, \pm 2, \dots$.
In typical log-periodic sequences, $p(n-1) \gg p(0)$, and if the signal were 
viewed as periodic, then there would be a large step between $p(n-1)$ and 
$p(n) = p(0)$. 
The step is primarily an artifact of the implicit DFT assumption
of periodicity, but
the large artificial step could dominate the behavior of the power spectrum.
As for windowing, typical window functions either leave a large boundary
step (e.g., rectangular window) or disregard the fact that the LPPL
behavior becomes more pronounced at the boundaries of the observation 
window (e.g., Hann).

In this paper, the discontinuity is obviated by {\em reflection\/}:
given a sequence $p(0), p(1), \dots, p(n-1)$, 
a new sequence is constructed by juxtaposing the original sequence
and a reverse version of the same sequence:
$p(n-1), \dots, p(1), p(0), p(1), \dots , p(n-2)$,
and assuming that the new sequence is periodic with period $2 (n - 1)$.
The reflected sequence has no abrupt discontinuity 
between the first and last element, and
thus no such step is visible in its spectrum.
We also tried alternative approaches, such as appending after $p(n-1)$
values that smoothly interpolate between $p(n-1)$ and $p(0)$ over a long
time frame, but we did not find significant differences between these 
alternatives and reflection.

To reiterate, given a finite sequence or a function defined over a finite
interval, the Fourier transform always requires 
an assumption of some sort on the sequence behavior
outside the given interval. For example, if the sequence is left unchanged,
the DFT implicitly assumes that the sequence is periodic.
In this light, reflection is a natural choice because, 
unlike implicit periodization or windowing, it eliminates abrupt 
discontinuities at the boundaries. 
Reflection also lead to a symmetrical analysis of bubbles
and anti-bubbles (i.e., a bubble-like behavior in reverse with
rapidly decreasing prices \cite{lppl-book}).

\begin{figure}
\begin{center}
\includegraphics[width=8cm]{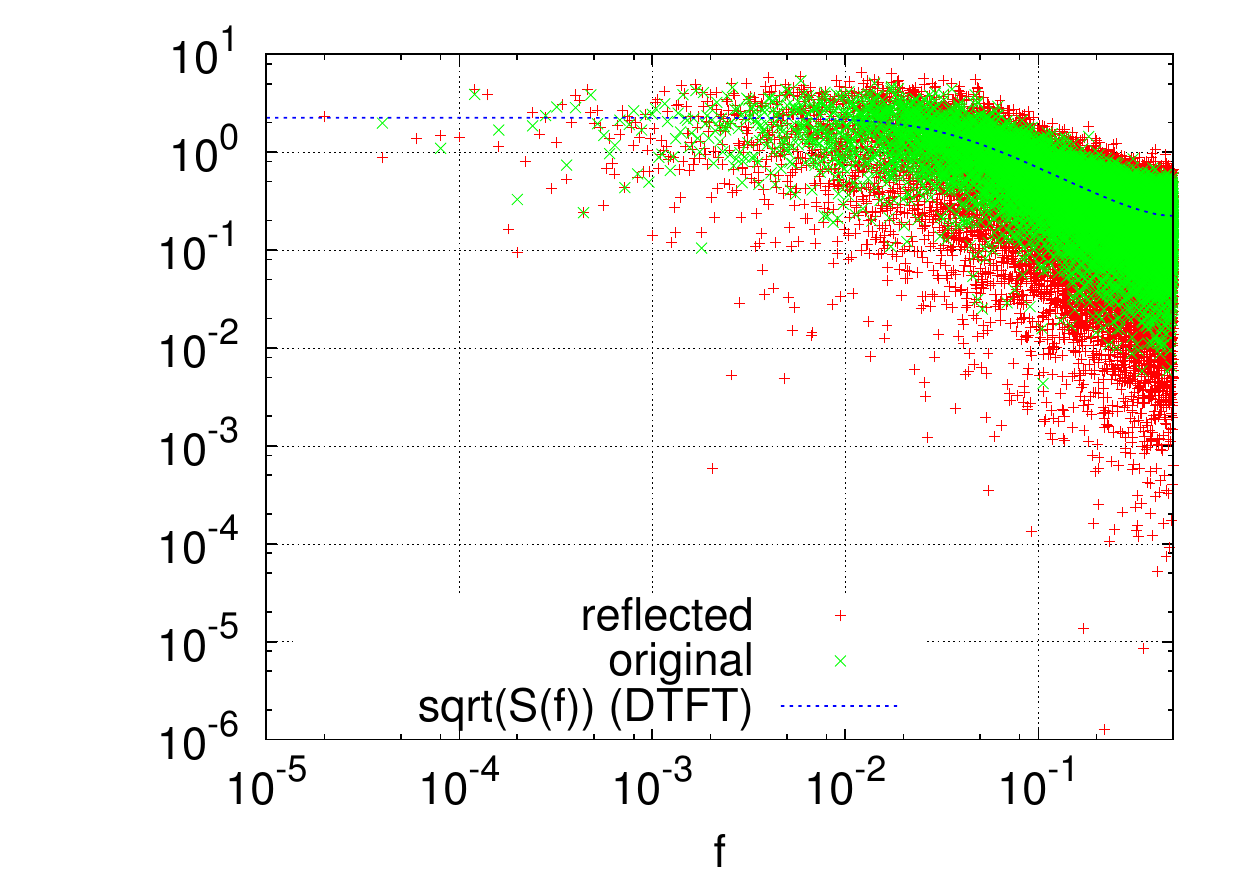}
\end{center}
\caption{DFT amplitude of a realization ($\sigma=.2$, $\tau=5$)
of the \ou process, of its reflection, and $\sqrt{S(f)}$.}
\label{fig:ou-reflect}
\end{figure}

If reflection is applied to a sequence $p = \ell + \nu$ 
that is the sum of a log-periodic component and a noise component, 
reflection will apply indiscriminately
to both components. Thus, it is possible in principle that reflection would
alter the power spectrum of noise. However, several
Monte Carlo simulations showed that reflection leaves the power spectrum 
of \ou noise practically unaffected.
For example, Figure \ref{fig:ou-reflect} gives
the DFT amplitude of a realization of an \ou process, 
the DFT amplitude of the reflection of the same process, and $\sqrt{S(f)}$. 
Reflection leaves the same general spectral behavior because, intuitively,
the autocorrelation of stochastic signals is invariant to time 
reversal and the autocorrelation of an \ou process decays exponentially.

\section{Log-Periodic Spectra}
\label{sec:lpplspectrum}
To derive the LPPL spectrum, 
we will break down the LPPL model into its two main components of
power law (without oscillations) and 
log-periodicity (without super-exponential growth).
Then, we will return to the general case with the insight gained
from the special cases, and verify numerically its spectral behavior.

\subsection{Power Law}
\label{sec:powerlaw}
Consider first the case when $C = 0$, i.e., the LPPL function is 
completely determined by its power law behavior.
If $m = C = 0$ (or if $B = 0$), $\ell$ reduces to a constant, and
its transform $L(f) = 0$ for $f \neq 0$.
If $C = 0, m = 1$, the reflected LPPL reduces to a triangular wave, for which
it is known that $|L(f_i)| \propto 1 / f_i^2$ when $i$ is odd
and $L(f_i) = 0$ when $i > 0$ is even.

The super-exponential behavior is the most pronounced when 
$m \rightarrow 0^+$, and $A, B$ change
as a function of $m$ so that $\ell$ is not a constant.
First, suppose for the time being that $T = n$
(the same result will hold for $T > n$ except for changes in the normalization
constants).
If $A$ is multiplied by a factor independent of $t$, 
$L(f)$ remains unchanged ($f \neq 0$).
Similarly, if $B$ is multiplied by a factor independent of $t$, 
$L(f)$ is multiplied by the same factor ($f \neq 0$).
In other words, a normalization of the constants $A$ and $B$ renormalizes
$L(f)$ but does not alter its qualitative behavior.
Assume that $A$ and $B$ are renormalized so that $\ell(0) = 0$
and $\int_0^T \ell(t) dt = 1 / 2$.
Thus,
$A = \hat{A}(m) = (1 + 1 / m) / (2 T)$ and
$B = \hat{B}(m) = A / T^m$.
Let $\ell_m(t) = \hat{A}(m) - \hat{B}(m) (T - t)^m$.
Note that $\lim_{m \rightarrow 0^+} \hat{A}(m) = \infty$
and that we take the transform of reflected signals.
Hence, $\lim_{m \rightarrow 0^+} \ell_m(t)$ is the delta function,
and its spectrum is a constant (specifically, $1 / 2 T$ with the 
common conventions on the DFT normalization factor).
\begin{figure}
\begin{center}
\includegraphics[width=8cm]{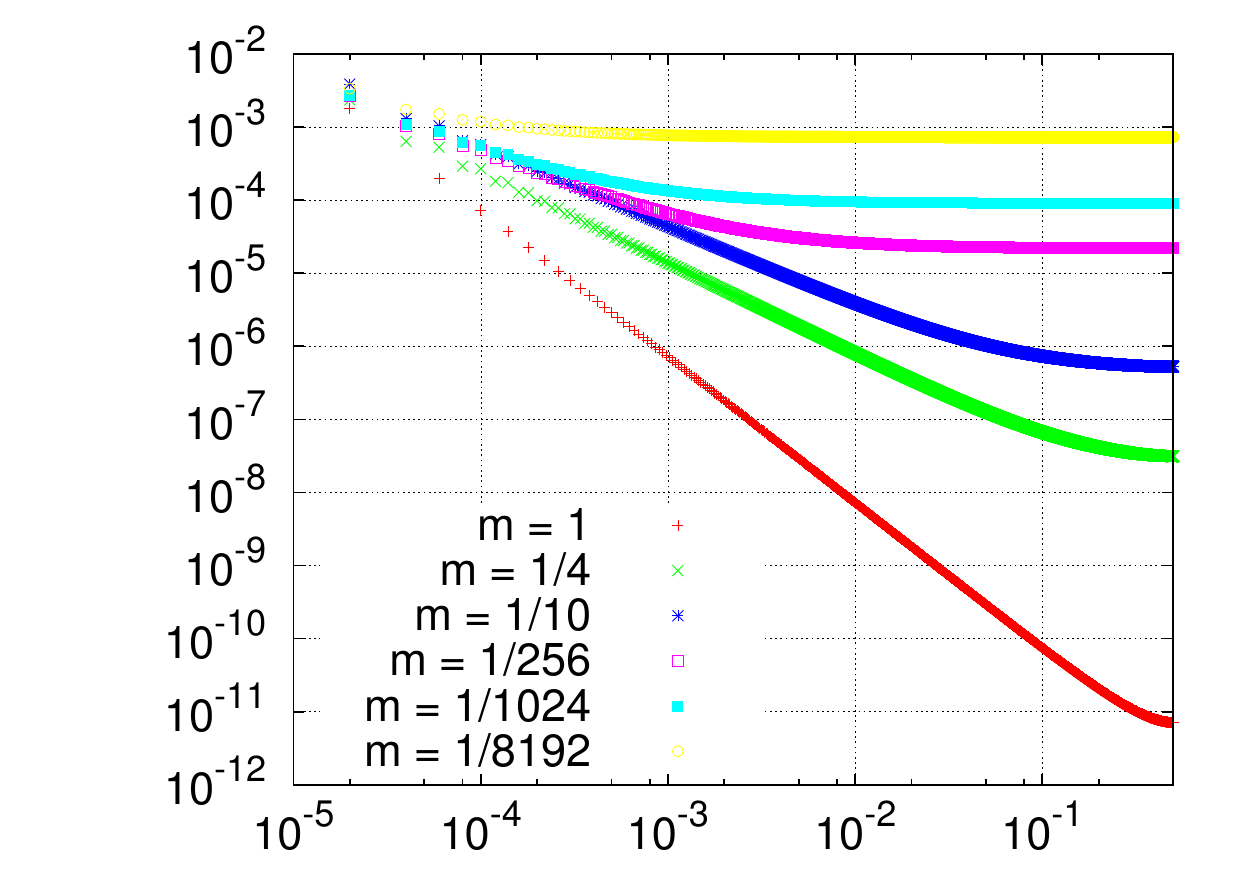}
\end{center}
\caption{DFT amplitude of the power law $\ell_m$.}
\label{fig:mddft}
\end{figure}
Figure \ref{fig:mddft} depicts the convergence of $\ell_m$ to the delta
function in the frequency domain.
When $m \neq 0$ is small, a flat spectrum characterizes the higher 
frequencies. 
As $m \rightarrow 0^+$, a constant spectrum is found in a progressively 
larger frequency interval.
Meanwhile, the spectrum at low frequencies has an increasingly flatter slope:
$|L(f)| \propto 1 / f^2$ at $m = 1$, 
$|L(f)| \propto 1 / f$ at $m \simeq 0.1$, and so on.
The amplitude $|L(f)|$ decreases faster than $1 / f$ for the range 
$m \geq 0.1$ that defines the ``stylized features of LPPL'' \cite{OU}.

\begin{figure}
\begin{center}
\includegraphics[width=8cm]{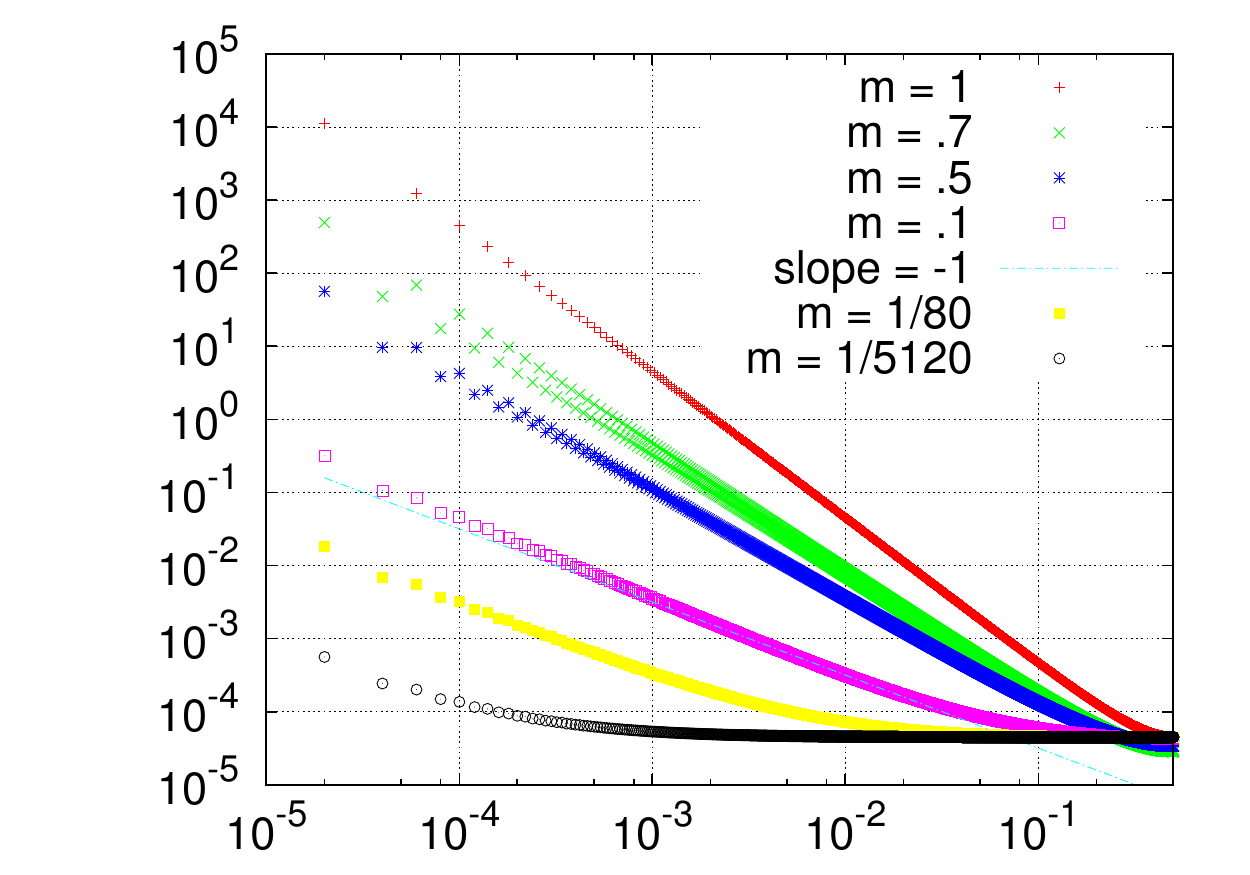}
\end{center}
\caption{DFT amplitude of an LPPL function for various values of $m$
($A = 10, B = 0.01, T = n = 25000, C = 0$).}
\label{fig:mdft}
\end{figure}

The same behavior holds also when $n < T$ by appropriately scaling
the normalization constants of $A$ and $B$.
The same results hold qualitatively for $A, B$ constant:
the spectrum is flat although the absolute values do change
depending on the normalization factor (see, for example, Figure \ref{fig:mdft}).

\subsection{Log-Periodicity}
\begin{figure}
\begin{center}
\includegraphics[width=8cm]{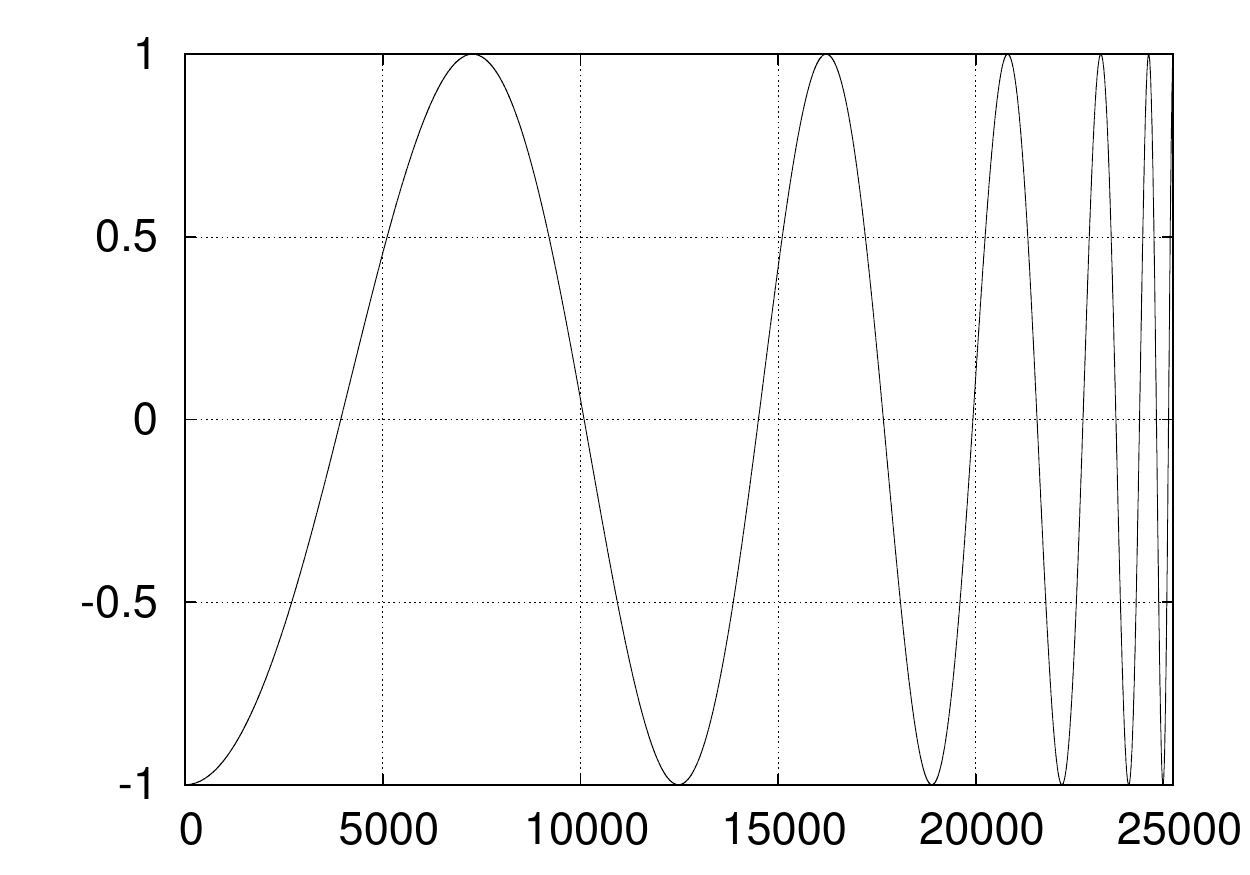}
\end{center}
\caption{An LPPL function that highlights the oscillatory term
as a function of non-linear time scaling
($A = 1$, $B = 1$, $C = -1$, $m = 0$, $\omega = 3 \pi$, 
$n = 25000$,
$T = n / (1 - e^{-11 \pi / \omega})$, $\phi = -\omega \ln (T - n)$).}
\label{fig:fm}
\end{figure}

\begin{figure}
\begin{center}
\includegraphics[width=8cm]{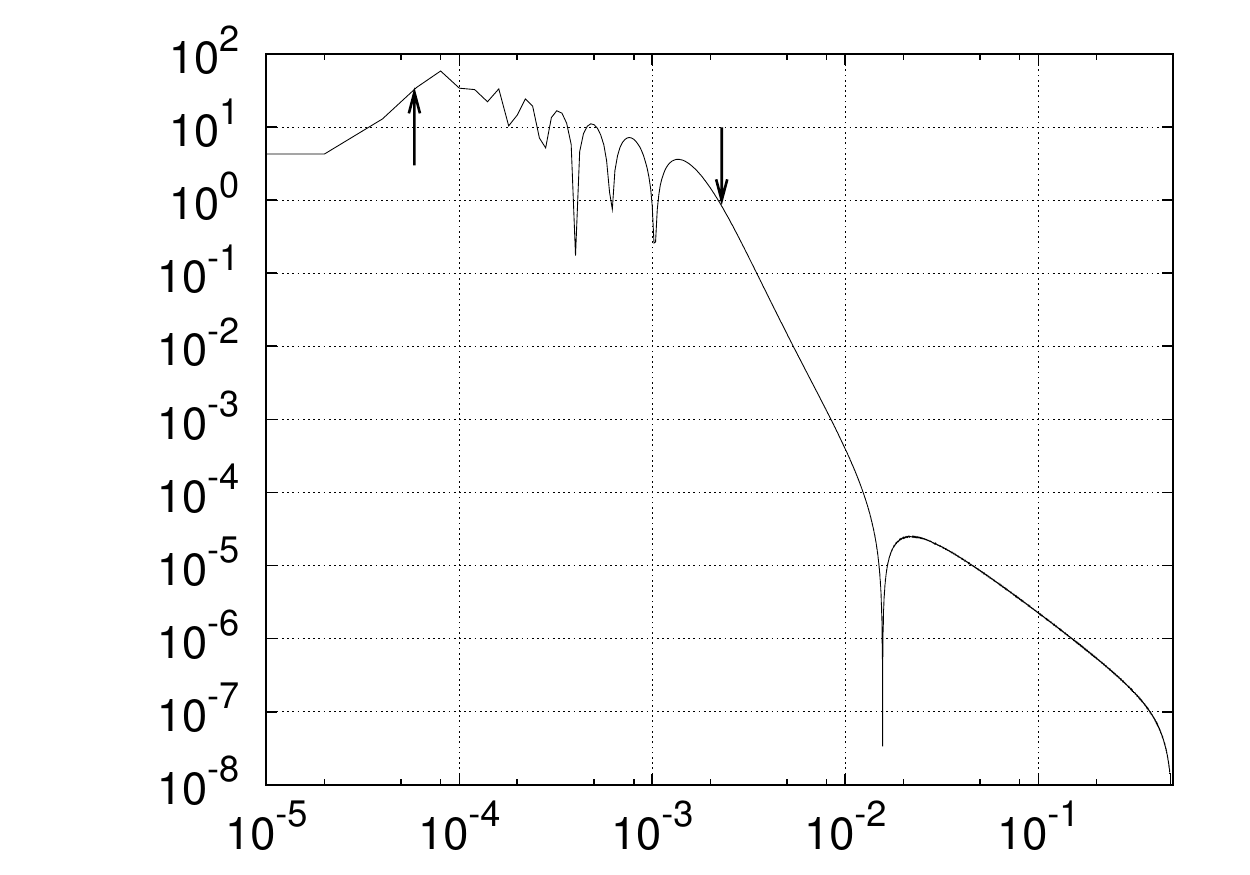}
\end{center}
\caption{The DFT amplitude of the LPPL function in Figure \ref{fig:fm},
with arrows pointing to the minimum and maximum frequencies
of the log-periodic spectrum.}
\label{fig:fmdft}
\end{figure}

If $m = 0$, then 
$\ell(t) = A - B (1 + C \cos(\omega \ln (T - t) + \phi))$
(see Figure \ref{fig:fm} for an example).
By discounting the constant terms, the resulting signal 
is $\tilde{\ell}(t) = \cos(\omega \ln (T - t) + \phi)$.
The signal $\tilde{\ell}(t)$ includes a non-linear time scaling, which
can be interpreted as the frequency modulation
of a carrier signal $\sin (2 \pi f(t) t + \alpha)$ with some underlying
signal $f(t)$.
Given a frequency modulated signal, such as $\tilde{\ell}(t)$, 
its power spectrum resides almost entirely in the band between the maximum and 
minimum value of the modulating signal \cite{carson}.
In practice, 
the differential frequency at time $t$ is given by the derivative of phase 
$\omega \ln (T - t) + \phi$ normalized by $2 \pi$, and is thus equal to
$\omega / (2 \pi (T - t))$.
Hence, the minimum frequency is $\omega / (2 \pi T)$ and the 
maximum frequency is $\omega / (2 \pi (T - n))$.
Figure \ref{fig:fmdft} shows an example along with the predicted frequency
bounds.

\begin{figure}
\begin{center}
\includegraphics[width=8cm]{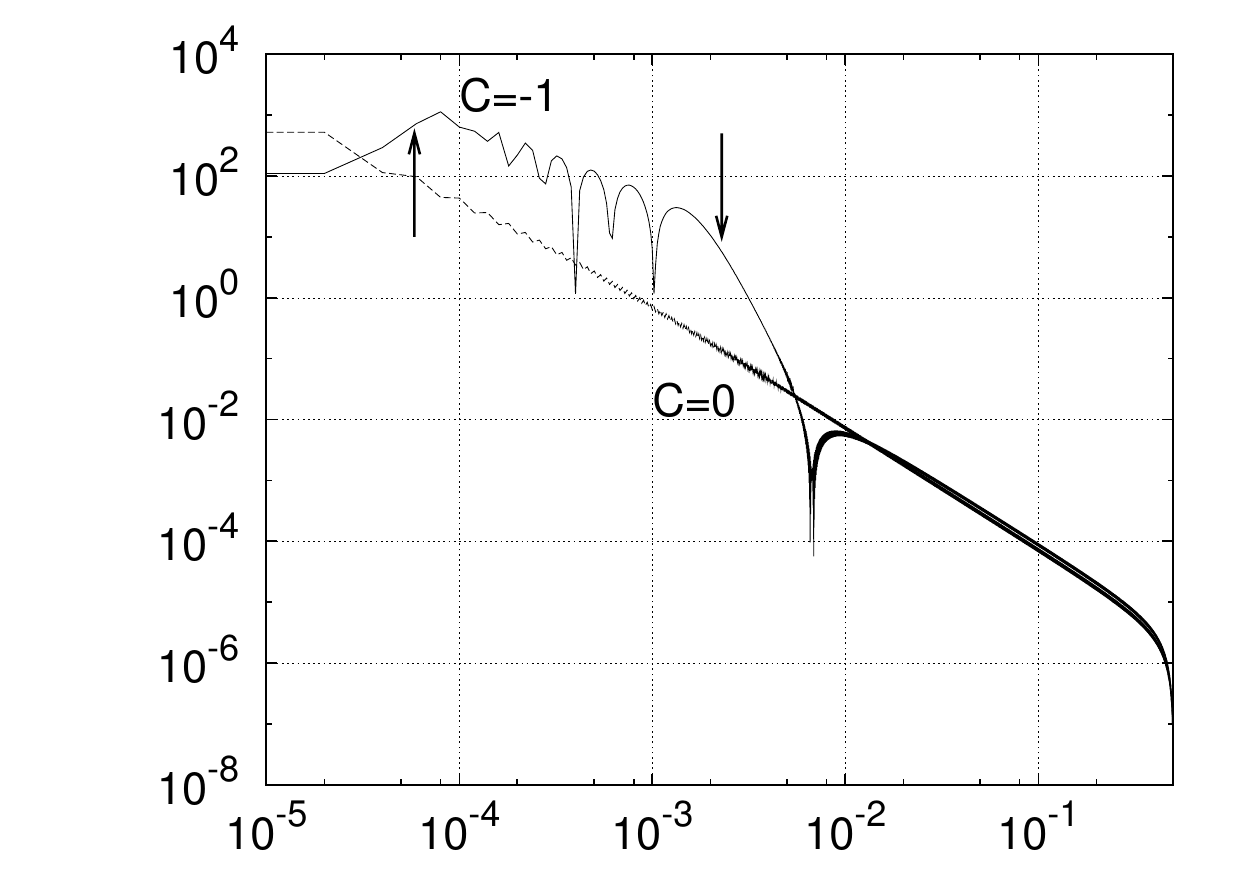}
\end{center}
\caption{The DFT amplitude of the LPPL function in Figure \ref{fig:fm} but
with $m = 0.3$,
with arrows pointing to the minimum and maximum frequencies 
of the log-periodic spectrum.}
\label{fig:fmmdft}
\end{figure}

Frequency-modulated signals are known to be resilient to changes in 
signal amplitude. In particular, the spectrum of
$\tilde{\ell}_m(t) = (T - t)^m \cos(\omega \ln (T - t) + \phi)$
should be qualitatively close to the spectrum of $\tilde{\ell}_0(t)$.
For example, Figure \ref{fig:fmmdft} shows the spectrum of
Figure \ref{fig:fmdft} except that now $m = 0.3$.
The maximum and minimum frequencies still delimit the area where the 
frequency modulated signal has predominant power.

Incidentally, the critical time $T$ is in a special position in that it
affects both power law (due to $(T - t)^m$ term) and log-periodicity 
(due to the $\ln (T - t)$ term).
Correspondingly, $T$ determines both the maximum frequency 
$\omega / (2 \pi (T - n))$ in the frequency modulated spectrum
and the normalization constants $\hat{A}(m), \hat{B}(m)$ in the 
power law spectrum.
In both cases, the effect of $T$ is similar in that, when $T$ approaches $n$,
the maximum frequency and the normalization constants both increase.

Frequency modulation analysis could be regarded as the dual of the 
Lomb transform of the detrended signal \cite{lppl-fit} in that
the frequency bounds are useful to estimate log-periodicity,
whereas Lomb analysis is useful to confirm it after the LPPL parameters 
have already been estimated.

\subsection{Combined LPPL}
The two LPPL hallmarks are power law and log-periodicity. 
Each translates directly into the frequency domain.
The power law implies a flat spectrum at high frequencies and
moderate slope at low frequencies.
Log-periodicity implies a bounded spectrum similar to that of a
frequency modulated signal.
Given the spectrum $|P|^2$ of prices during a bubble,
we would expect to find either (or both) signatures
of LPPL in the frequency domain.

\begin{figure}
\begin{center}
\includegraphics[width=8cm]{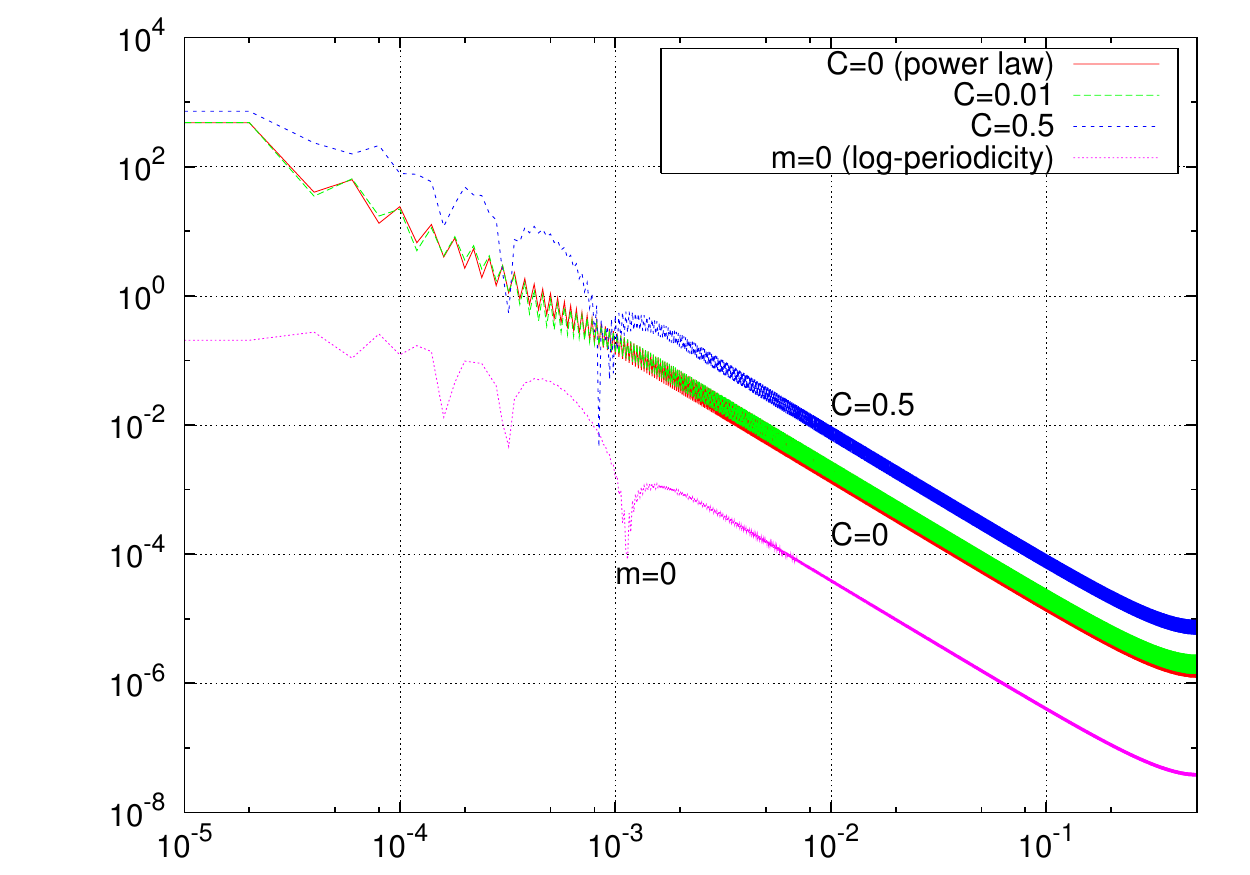}
\end{center}
\caption{$|L(f)|$ for 
$A = 100, B = 0.01, T = 26000, n = 25000, \omega = 2 \pi, \phi = 0$
in the cases of pure power law ($m = 0.7, C = 0$),
pure log-periodicity ($m = 0, C = 0.5$),
and two intermediate cases ($m = 0.7, C = 0.01, 0.5$).}
\label{fig:Cdft}
\end{figure}

\begin{figure}
\begin{center}
\includegraphics[width=8cm]{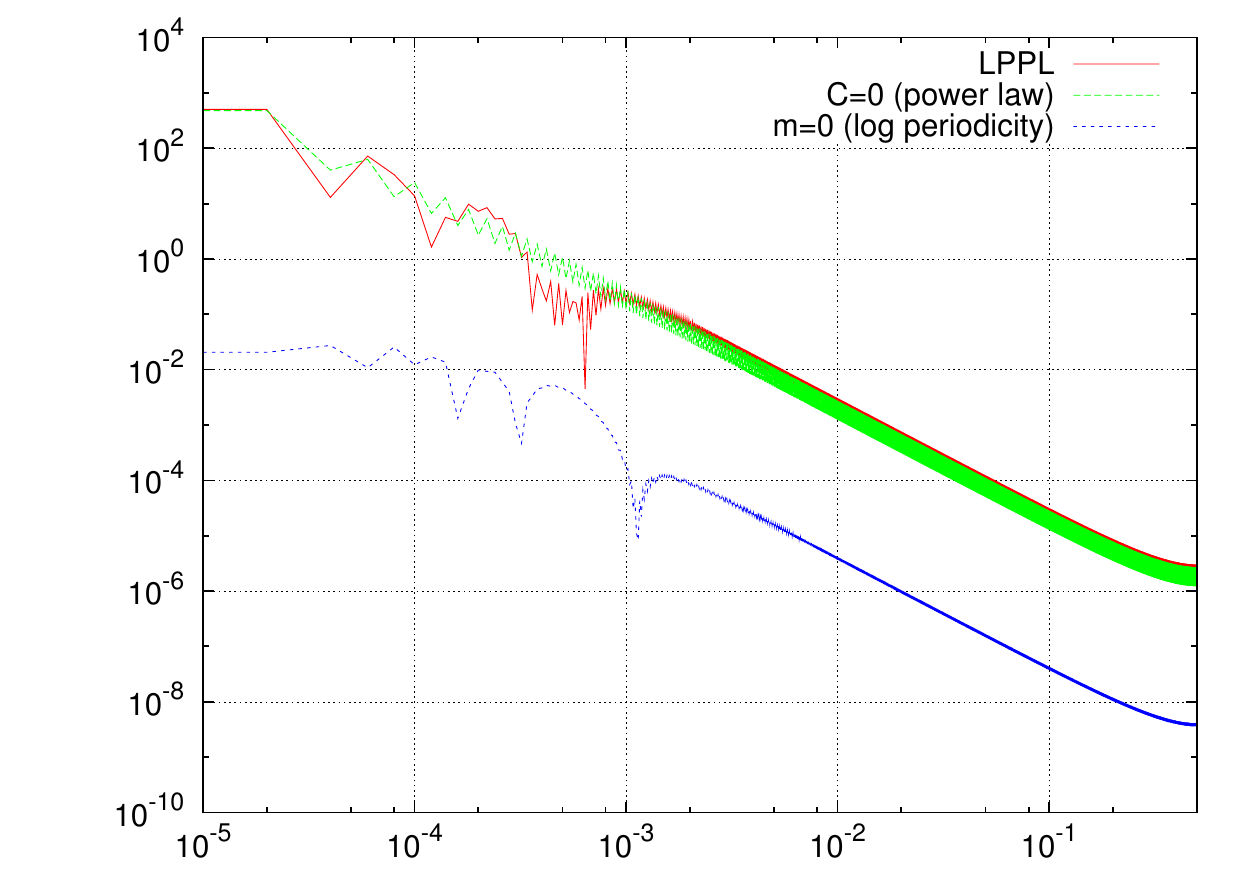}
\end{center}
\caption{$|L(f)|$ for 
$A = 100, B = 0.01, T = 26000, n = 25000, \omega = 2 \pi, \phi = 0$
in the cases of pure power law ($m = 0.7, C = 0$),
pure log-periodicity ($m = 0, C = 0.05$),
and combined case ($m = 0.7, C = 0.05$).}
\label{fig:Csmalldft}
\end{figure}

The general LPPL spectrum is analytically intractable,
and it was estimated numerically.
Sequences were generated for hundreds of different LPPL parameter sets,
the respective spectra were calculated numerically and plotted.
In general, to make sense of the spectrum, it is helpful to interpret it
qualitatively in terms of the two main features of 
power law and log-periodicity.
Specifically, a LPPL spectrum can be thought as the superposition of 
a power law spectrum and a log-periodic spectrum, where the superposition
is weighted by the parameter $C$. 
For example, Figure \ref{fig:fmmdft} compares an LPPL spectrum with that
of a pure power law with the same parameter values (but $C = 0$).
As another example, Figure \ref{fig:Cdft} shows the spectrum for 
a pure power law function ($m = 0.7, C = 0$), 
a pure log-periodic function ($m = 0, C = 0.5$),
and two intermediate cases ($m = 0.7, C = 0.01, 0.05$).
The intermediate cases have a spectrum that is visually the superposition
of the two extreme spectra. Furthermore, higher values of $C$ make the 
log-periodic component more visible.
If $C$ is small, the frequency modulation lobes are typically visible around
the trend rather than above it (e.g., Figure \ref{fig:Csmalldft}).
Results were similar for all other combinations of parameters, and omitted.
Although it is helpful to comprehend a general LPPL spectrum as a superposition
of feature, the intuition can seldom be translated into a closed expression.
However, it is always useful to think the LPPL spectrum as the combination 
of two broad features.
In particular, given a noisy LPPL, we would expect to find the two LPPL
signatures in the frequency domain.

\section{De-Noising}
\label{sec:de-noise}
We now turn to discuss the extent to which it is possible
to extract an underlying LPPL signal from noisy price measurements.
The expectation would intuitively be that the mean-reverting noise smooths out 
relatively quickly so that the underlying LPPL behavior should be visible at 
the lower frequencies.
More precisely, the potential for de-noising depends on achieving high
values of the 
{\em signal-to-noise ratio\/} $R(f) = |L(f)|^2 / S(f)$.
The signal-to-noise ratio $R(f)$ is related to the non-causal
{\em Weiner filter\/} $K(f) = 1 / (1 + 1 / R(f))$.
The Weiner filter $K$ has the property that the filtered signal
$\hat{\ell}(i) = \sum_{j = -\infty}^{\infty} k(j) p(i-j)$
minimizes the mean square error $\sum (\ell(i) - \hat{\ell}(i))^2$
among all linear filters,
where the $k$'s are the inverse transform of $K$
and $p$ is the reflected and periodicized time series of prices.
If $R(f)$ decreases with $f$, then $K(f)$ is a low pass filter and,
intuitively, the filter replace each price $p(i)$ with a smoothed combination
of its neighbors so as to reject the noise.
The signal-to-noise ratio is a function of frequency and 
can take various shapes depending on the 
parameters of the underlying LPPL $\ell$ and \ou noise.

In the stylized LPPL, $m \geq 0.1$, and, with a pure power law ($C = 0$), 
$|L|^2$ always decreases faster than the slope of $S$.
Hence, the power law contributes to de-noising only if $|L(f)|^2 > S(f)$
at low frequencies.
In pure log-periodicity ($m = 0$), $|L(f)|$ is significant only
for frequencies $f \leq \omega / (2 \pi (T - n))$.
Again, log-periodicity contributes to de-noising only if $|L(f)|^2 > S(f)$
at low frequencies.

\begin{figure}
\begin{center}
\includegraphics[width=8cm]{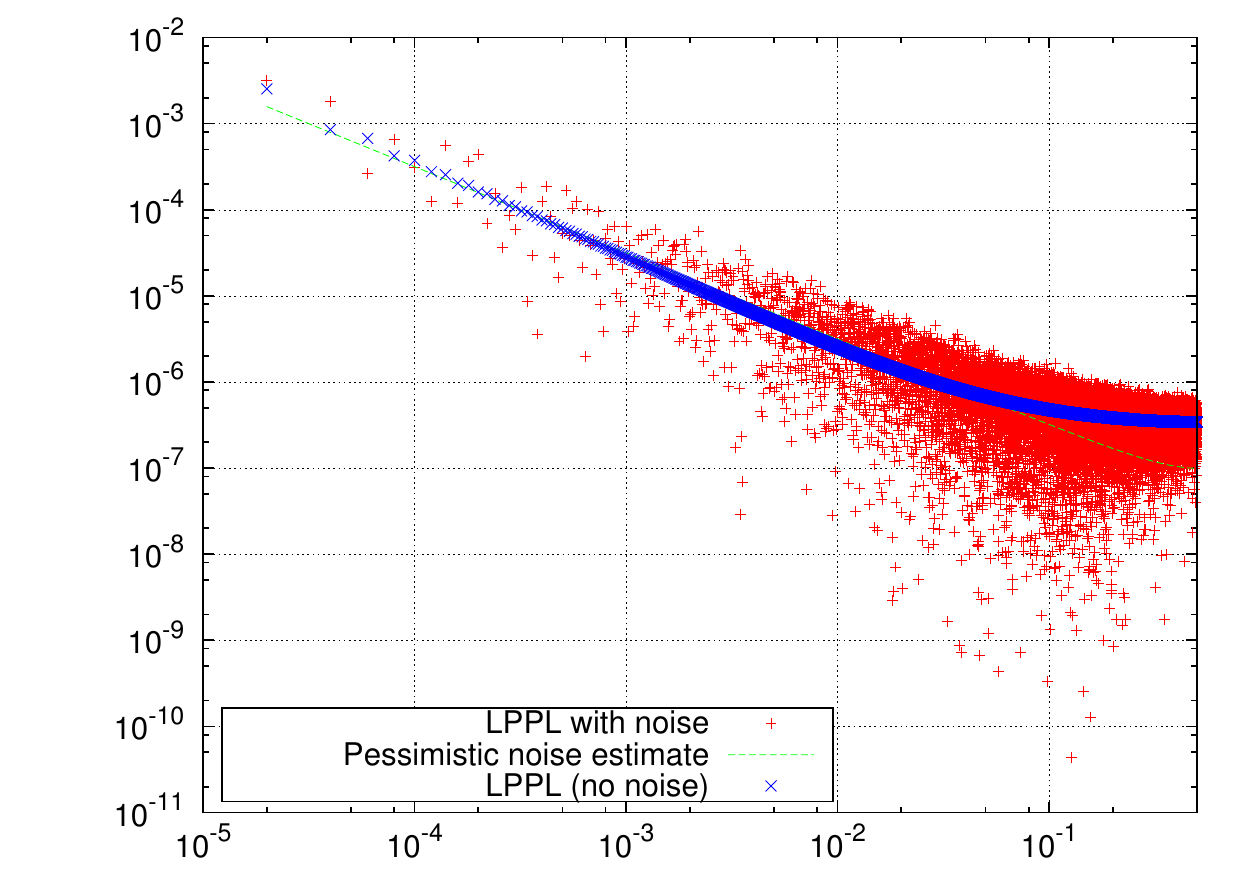}
\end{center}
\caption{Spectrum $|L(f)|$ of $\ell_{0.1}$ with and without noise,
and pessimistic noise estimate.}
\label{fig:critical}
\end{figure}

It is useful to examine the special case when 
$|L(f)| \propto 1 / f$ at low frequencies since $|L|$'s slope matches exactly
the slope of $S_{\infty}$.
In this case, it may be hard to ascertain the presence of a signal $\ell$
if Weiner noise is also present
since the LPPL spectrum almost perfectly overlaps with the noise spectrum.
Figure \ref{fig:critical} shows the amplitude $|L|$ of an LPPL $\ell_{0.1}$ 
without noise.
Then, Gaussian white noise was generated with variance $\sigma^2$ chosen 
so that the corresponding spectrum $S_{\infty}$ matches closely $|L|^2$
at low frequencies. 
The amplitude of the resulting price transform is shown 
in Figure \ref{fig:critical}
along with the pessimistic estimate defined in Section \ref{sec:pessimistic}.
The pessimistic estimate overlapped almost exactly the true noise spectrum
(whereas the maximum likelihood estimate did not produce finite parameter
estimates). 
The spectra $|L|^2$, $|P|^2$, and $S_{\infty}$ were practically identical
at low frequencies, but at high frequencies $|P|$ followed the larger $|L|$.
We conclude that even in the special case when the LPPL is hidden under
the noise because $m$ makes
$|L(f)| \propto 1 / f$ and 
$\sigma^2$ makes $|L(f)|^2 \simeq S_{\infty}$ at low frequencies, 
LPPL and noise can be distinguished by extracting a pessimistic noise 
estimate.
However, in this case, the difference is noticeable at high frequencies.

In summary, the intuition was that LPPL should be visible at the low 
frequencies, and it is correct in most cases.
Then, de-noising should be possible when the spectrum looks qualitatively
as in Figure \ref{fig:pessimistic}, with the price spectrum clearly above
the noise estimate for a significant range of frequencies at the low end
of the spectrum.
However, we found special combinations of parameters
(e.g., Figure \ref{fig:critical}) where the 
presence of LPPL is visible only from the highest frequencies.

If the time horizon $n$ approaches the critical time $T$, 
intuition would suggest that it should be easier to detect an LPPL trend.
Indeed, if the critical time $T$ is approaching, then 
the maximum log-periodic frequency is higher 
resulting into a wider frequency modulated spectrum,
and the normalization constant $\hat{B}$ is larger, 
shifting the spectrum upward.
In other words, when the critical time approaches, 
the LPPL function resembles more closely an impulse and 
additionally its oscillations are wilder, so that it is hard for 
mean-reverting noise to disguise the underlying LPPL trend.

\section{Evaluation}
\label{sec:evaluation}
Data sets are daily closing values of the major indexes and 
securities shown in Table \ref{table:dataset}.
The first four data sets are associated with major historical bubbles 
in the U.S. stock market. 
The last data set is a recent bubble in gold prices that burst in 
November 2009 \cite{sornette-gamble}.
These prices sequences stop closely before the time 
at which the price reached its maximum level during the bubble episode.
Therefore, the detection of an underlying LPPL trend should be relatively 
easy.
We have additionally considered individual stocks.
For example, we have investigated various tech stocks during the 1998-2000 
bubble and the Netflix stock in the 2005-10 period.
The results are qualitatively similar to those reported here, and omitted.
\begin{table}
\begin{center}
\begin{tabular}{|l|l|l|} \hline
{\bf Series\/} & {\bf Period\/} & {\bf Data Points\/} \\ \hline\hline
Dow Jones Industrial Average & June 1921-July 1929 & 2440 \\ \hline
S\&P 500  & July 1985--July 1987 & 527 \\ \hline
NASDAQ Composite & January 1994--February 2000 & 1555 \\ \hline
S\&P 500  & July 2003--June 2007     & 1000 \\ \hline
GLD       & March 2009--October 2009 & 171 \\ \hline
\end{tabular}
\end{center}
\caption{Price time series used in the evaluation.}
\label{table:dataset}
\end{table}

The maximum likelihood and pessimistic noise estimates were close to each other,
providing evidence that mean reversion is weak.
Both estimates fit well the price spectrum,
a sign that prices can be explained primarily as a Weiner process.
In most cases, the spectrum deviates from the pessimistic estimate only at
$f = 0, 1 / (2 n)$, which is consistent with adding Weiner noise to
deterministic exponential price trajectory.

\begin{figure}
\begin{center}
\includegraphics[width=8cm]{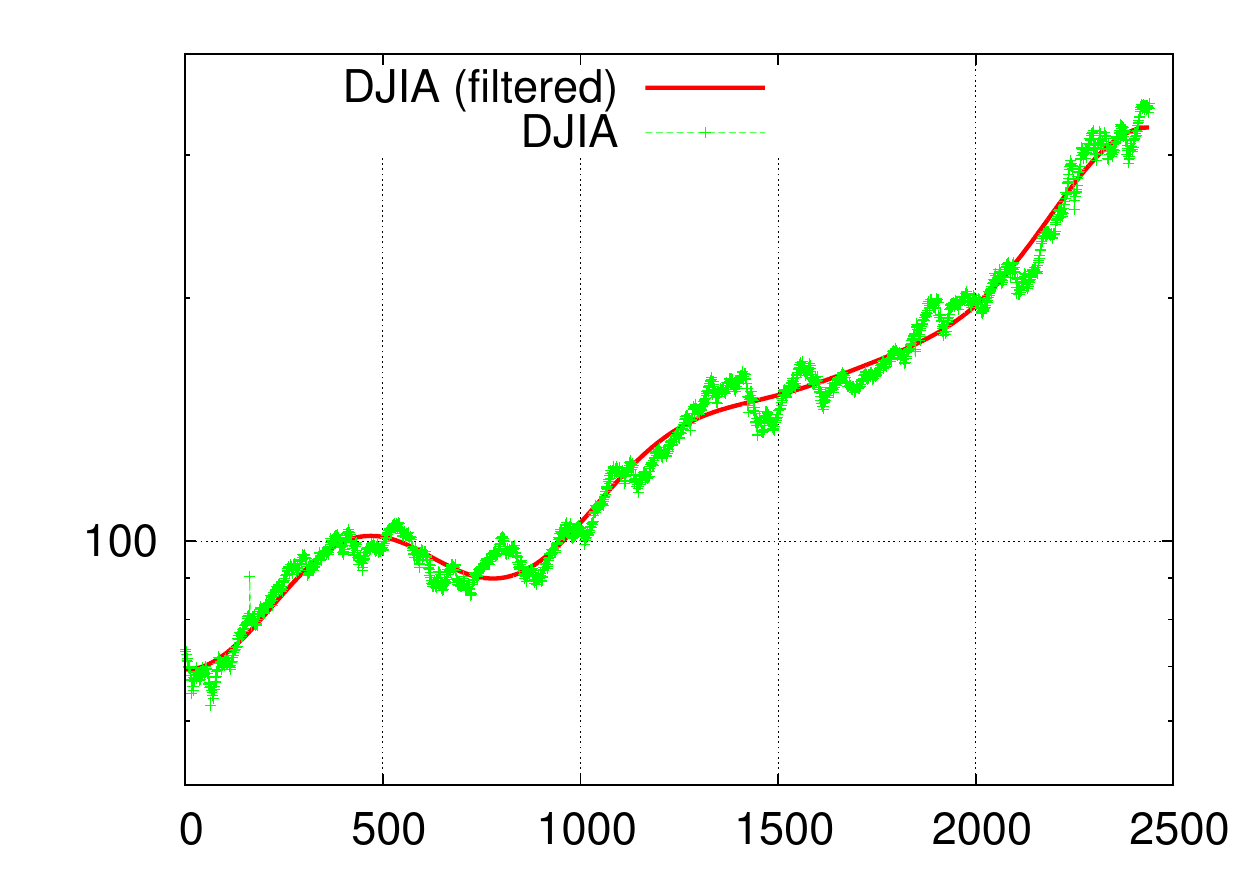}
\end{center}
\caption{Dow Jones 1929 with filtered price sequence ($\tilde{f} = f_9$).}
\label{fig:djia29}
\end{figure}

\begin{figure}
\begin{center}
\includegraphics[width=8cm]{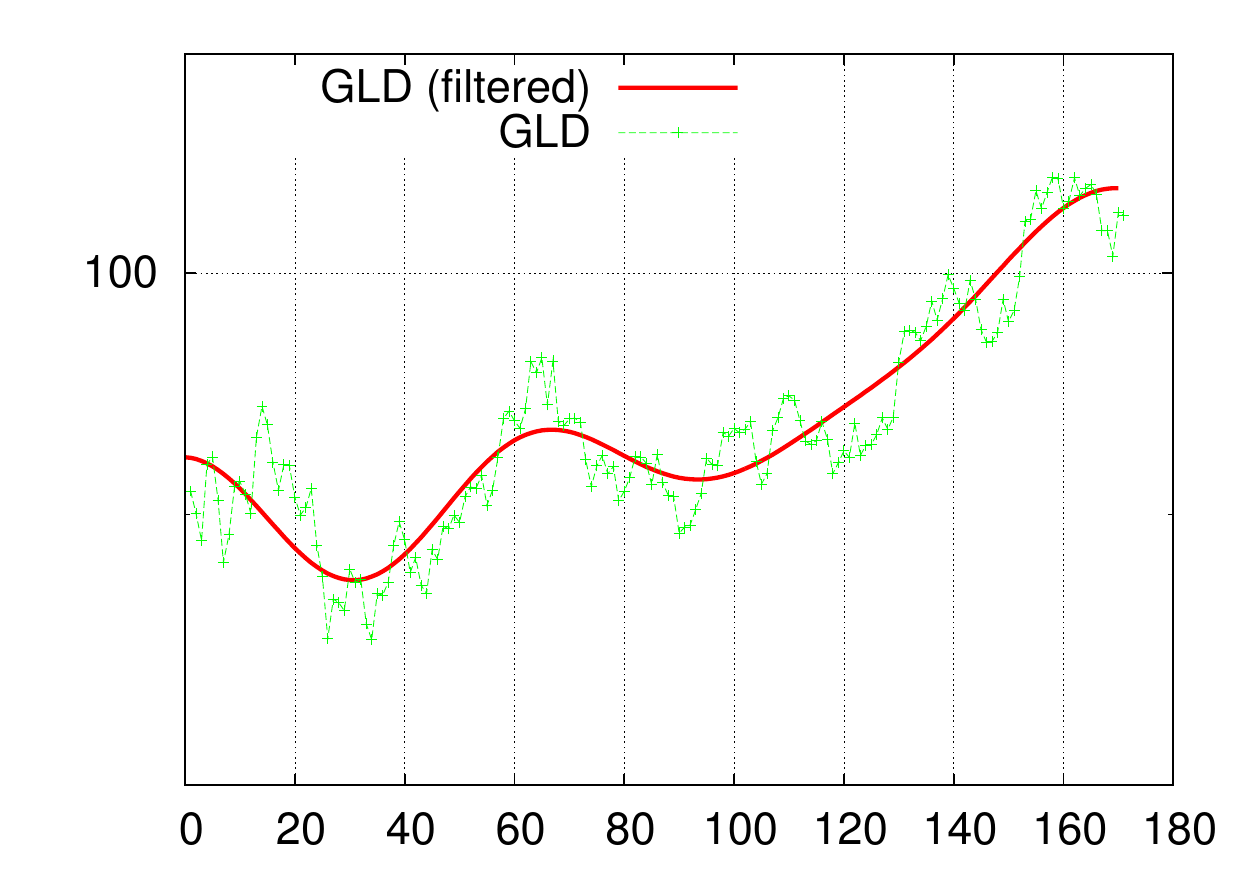}
\end{center}
\caption{Gold 2009 with filtered price sequence ($\tilde{f} = f_6$).}
\label{fig:gld09}
\end{figure}

As for the maximum likelihood estimate, the diffusion constant 
is relatively large.
The maximum likelihood estimate was slightly but consistently smaller 
at low frequencies than the price spectrum in two cases: 
Dow Jones 1929 and gold 2009. 
The discrepancy was investigated with a filter to remove the maximum likelihood
noise. 
The filter is non-causal and produces a signal $\tilde{\ell}$ with 
the property that $\tilde{L}(f) = L(f)$ if $f < \tilde{f}$ and 
$\tilde{L}(f) = 0$ otherwise, where $\tilde{f}$ is the smallest frequency $f$
at which $|L(f)|^2 < S(f)$.
The intuition is that the underlying LPPL spectrum $|L|^2$ decreases faster 
than the noise spectrum $S$, so if one takes the smallest frequency $f$ at 
which $|L(f)|^2 \geq S(f)$, the resulting low-pass filter
approximates a Weiner filter.
Figures \ref{fig:djia29} and \ref{fig:gld09}
show the original and filtered series.
Both filtered series show significant oscillations at the beginning
and slight super-exponential growth throughout.
However, the oscillations are absent toward the end of the filtered sequence.
Since log-periodic oscillations become progressively more rapid toward the
critical time,
even if log-periodicity were present in these price sequences, 
oscillations would be blocked by the filter at the highest log-periodic 
frequencies.
However, the filter only blocks frequencies when noise dominate the signal,
and so we conclude that log-periodicity, if at all present, has been disrupted
by noise.
Similarly, super-exponential growth is visually unclear both in the time and 
in the frequency domain.

In summary, noise had at best weak mean reversion and relatively high
variance, so that, even if LPPL underlied price dynamics, it was obscured
by noise.

\begin{figure}
\begin{center}
\includegraphics[width=8cm]{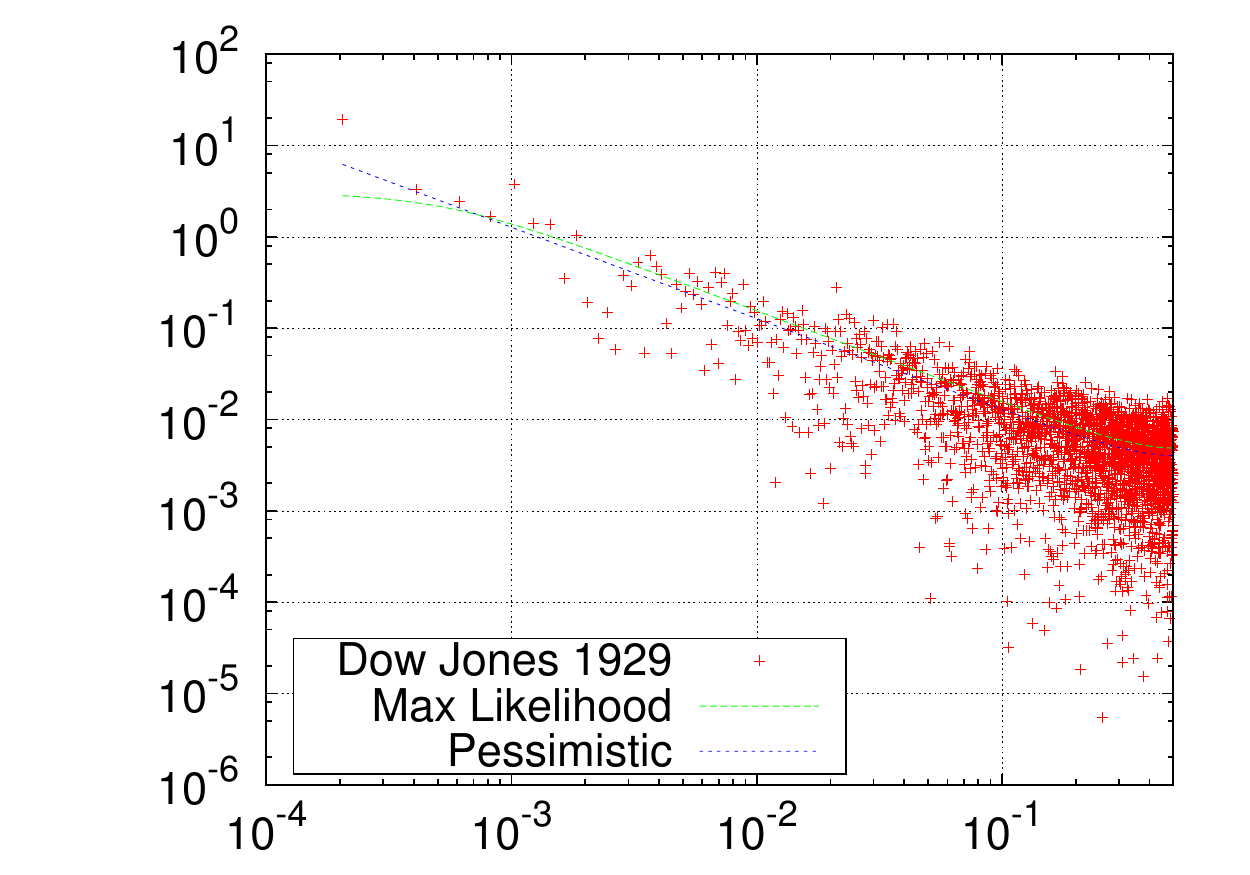}
\end{center}
\caption{Spectrum of the Dow Jones 1929 price sequence.}
\label{fig:djiadft}
\end{figure}
\begin{figure}
\begin{center}
\includegraphics[width=8cm]{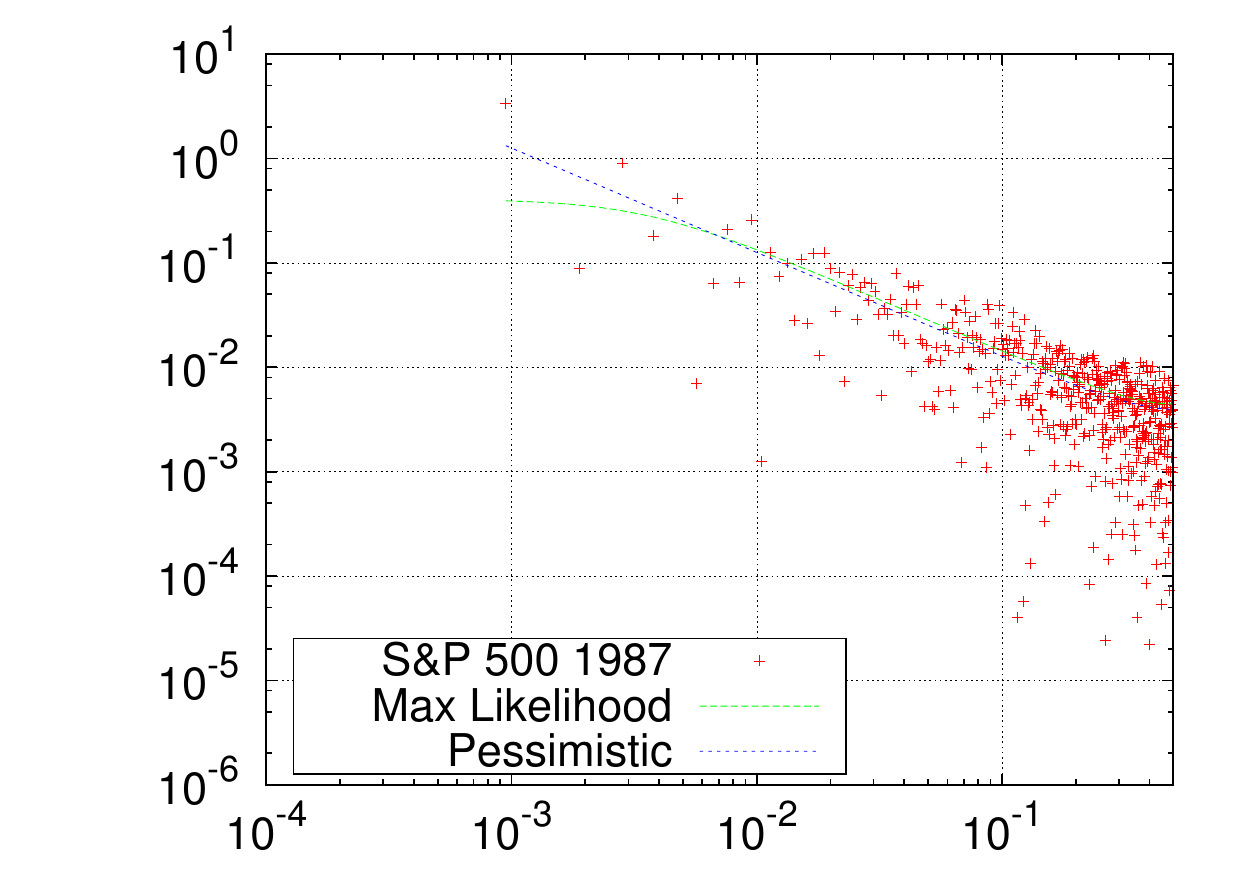}
\end{center}
\caption{Spectrum of the S\&P 500 1987 price sequence.}
\label{fig:gspc87dft}
\end{figure}
\begin{figure}
\begin{center}
\includegraphics[width=8cm]{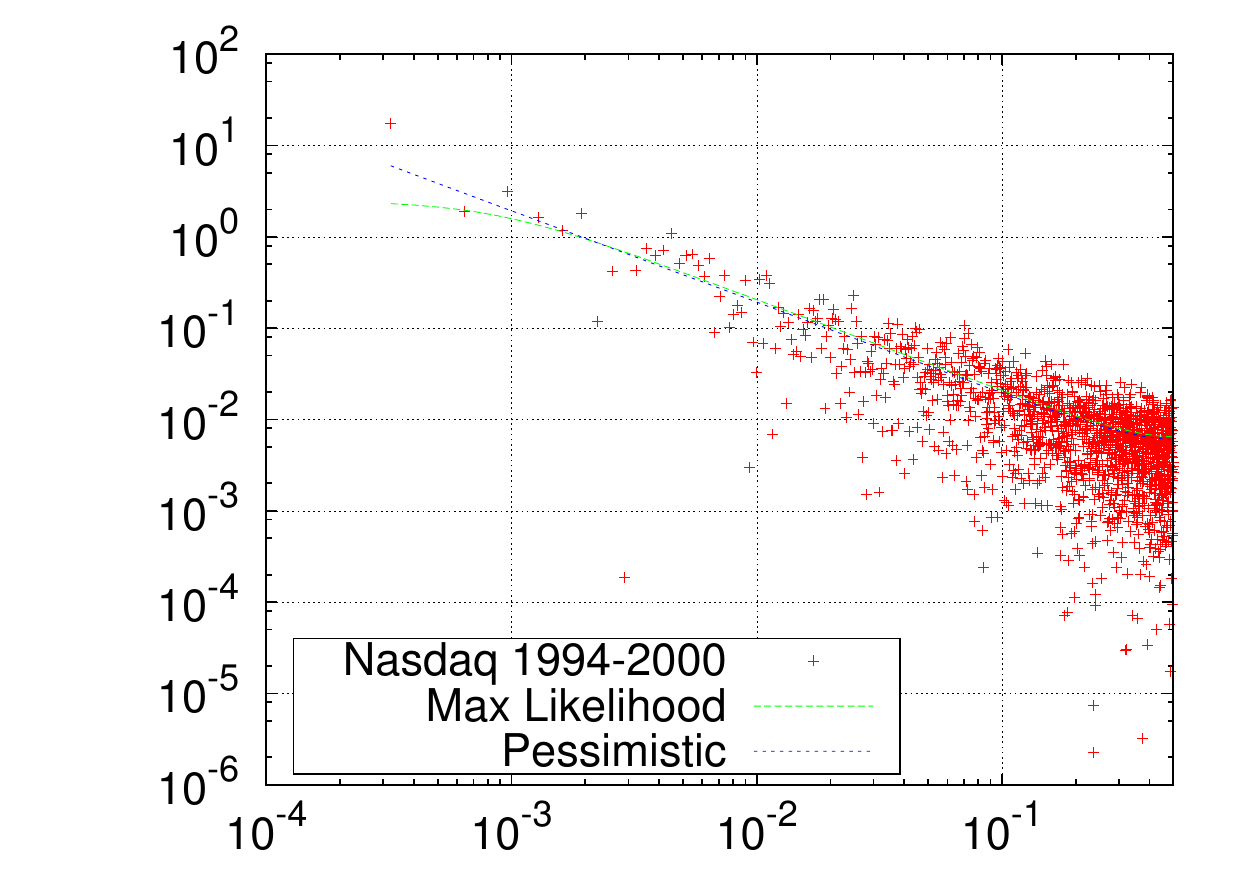}
\end{center}
\caption{Spectrum of the NASDAQ Composite 1994-2000 price sequence.}
\label{fig:ixic9400dft}
\end{figure}
\begin{figure}
\begin{center}
\includegraphics[width=8cm]{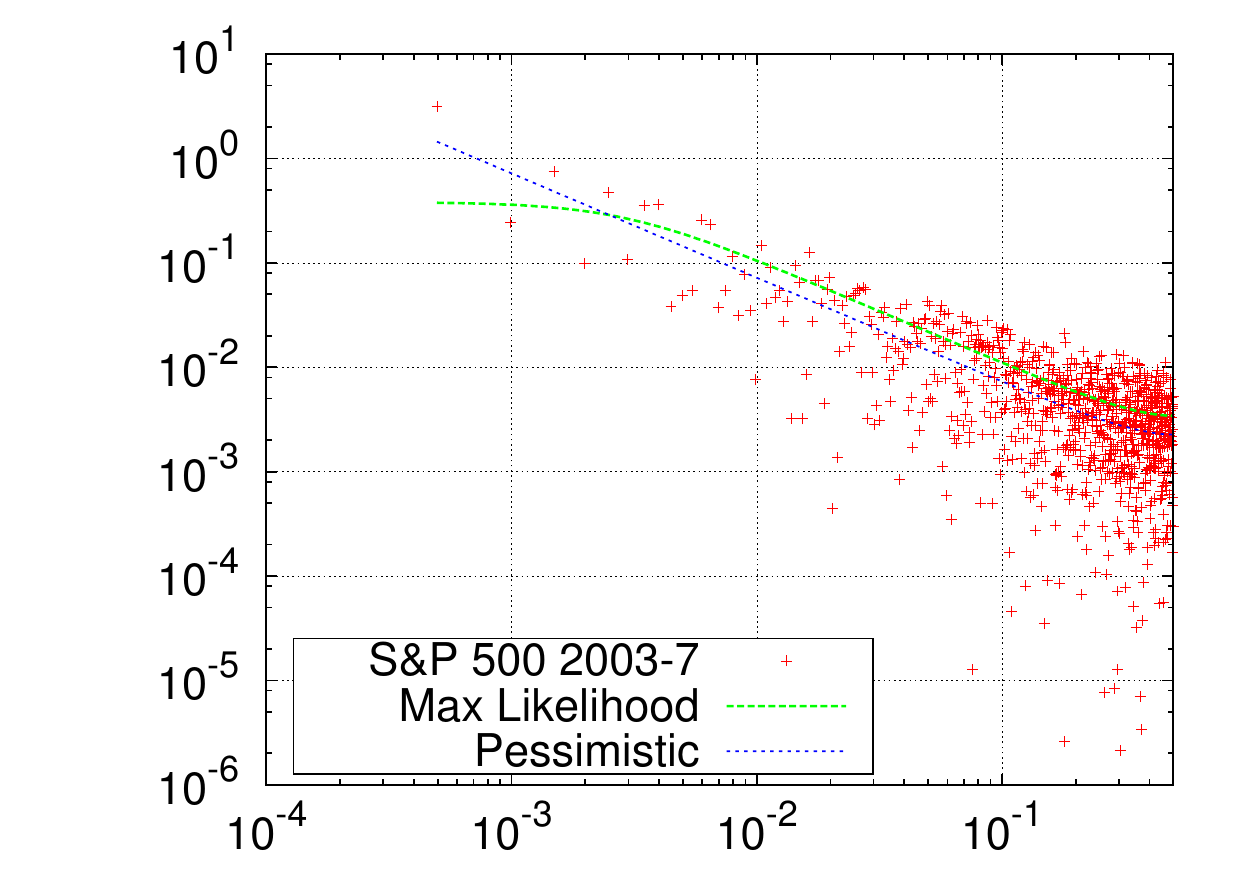}
\end{center}
\caption{Spectrum of the S\&P 500 2003-07 price sequence.}
\label{fig:gspc0307dft}
\end{figure}
\begin{figure}
\begin{center}
\includegraphics[width=8cm]{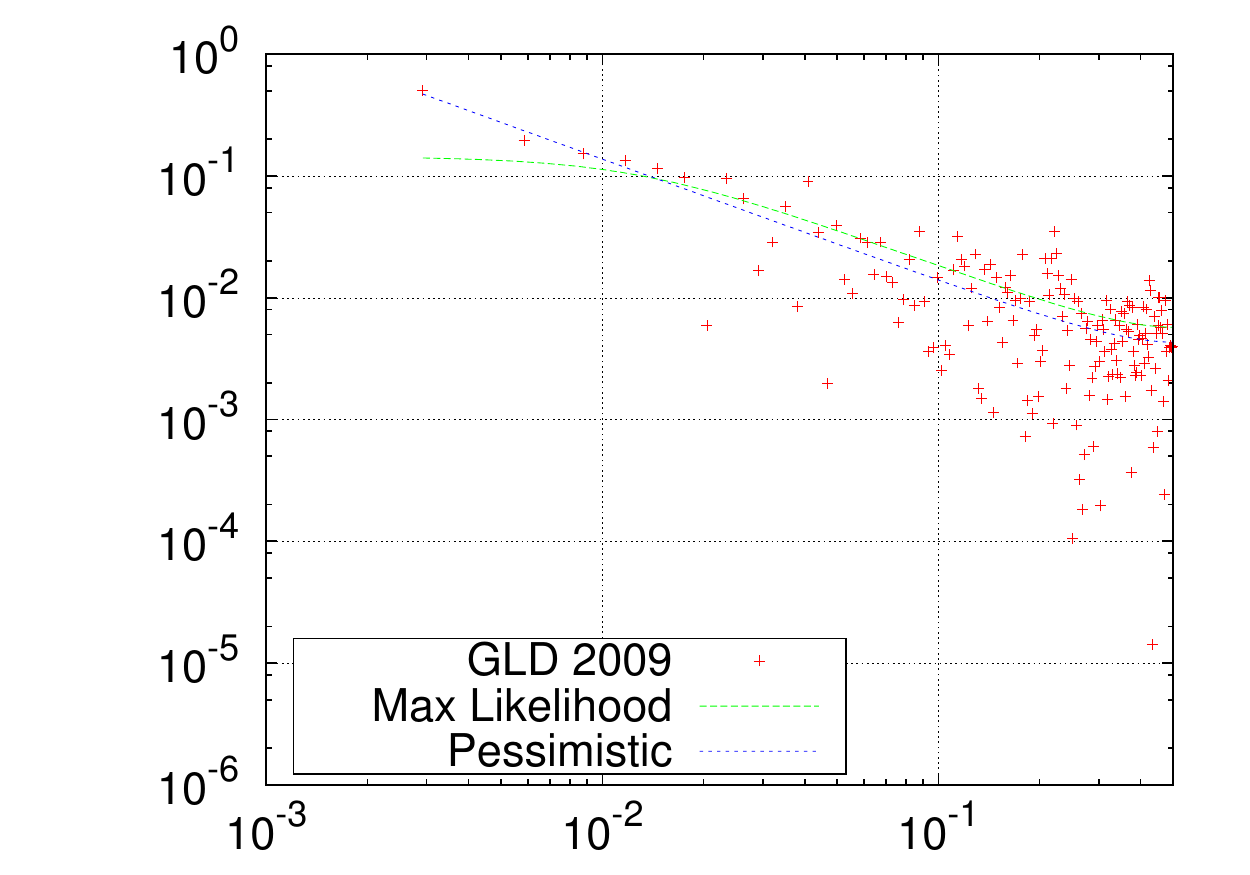}
\end{center}
\caption{Spectrum of the GLD (gold) 2009 price sequence.}
\label{fig:glddft}
\end{figure}

\section{Conclusions}
\label{sec:conclusion}
The estimators show that mean reversion is weak, and \ou noise is close
to a non-mean-reverting Weiner process. 
Furthermore, the signal-to-noise ratio $R(f)$ is low, and 
it mostly makes it possible to reconstruct an underlying 
exponential trend.
Even in those cases where $R(f) > 0$ for a range of low frequencies,
noise makes it impossible to reconstruct either log-periodicity or 
a power law.
In short, due to noise, there is no trace of LPPL during bubbles.

However, it is possible in principle that LPPL underlies price dynamics, 
but that it cannot be isolated with pure frequency methods. 
Future work will investigate different transforms, such as wavelets, that
may make it possible to discern explosive bubble growth.
Furthermore, if the underlying LPPL were expressed as a dynamic system with 
known parameters, it would be possible to use the additional information in
de-noising filters that extend a pure frequency approach.

\bibliographystyle{apalike}
\bibliography{lppl}

\end{document}